\documentclass[twocolumn,showpacs,preprintnumbers,amsmath,amssymb,aps,pra]{revtex4-1}

\usepackage{dcolumn}
\usepackage{graphicx}
\usepackage{subfigure}
\usepackage{subfiles}
\usepackage{natbib}
\usepackage{soul}
\usepackage{float}
\usepackage{bm}
\usepackage{multirow}
\usepackage[normalem]{ulem}
\usepackage[usenames,dvipsnames]{color}
\usepackage{xr-hyper}
\usepackage{hyperref}
\newcommand{\beq}{\begin{equation}}
\newcommand{\eeq}{\end{equation}}
\newcommand{\barr}{\begin{eqnarray}}
\newcommand{\earr}{\end{eqnarray}}

\begin{document}

\title{
Energetics and cathode voltages of LiMPO$_4$ olivines (M = Fe, Mn)
from extended Hubbard functionals
}

\author{Matteo Cococcioni} 
\thanks{Current address: Department of Physics, University of Pavia, Via Bassi 6, 27100 Pavia, Italy}

\author{Nicola Marzari} 

\affiliation{Theory and Simulation of Materials (THEOS), and 
National Centre for Computational Design and Discovery of 
Novel Materials (MARVEL), 
\'Ecole Polytechnique F\'ed\'erale de Lausanne (EPFL), 
CH-1015 Lausanne, Switzerland}

\date{\today}

\begin{abstract}
Transition-metal compounds pose serious challenges to
first-principles calculations based on density-functional theory (DFT),
due to the inability of most approximate exchange-correlation functionals to capture
the localization of valence electrons on their $d$ states, essential for a predictive modeling of their
properties.
In this work we focus on two representatives of a well known family of cathode materials
for Li-ion batteries, namely the orthorhombic LiMPO$_4$ olivines (M = Fe, Mn). 
We show that extended Hubbard functionals 
with on-site ($U$) and inter-site ($V$) interactions (so called DFT+U+V) can predict 
the electronic structure of the mixed-valence phases,
the 
formation energy of the materials with intermediate Li contents,
and the overall average voltage of the battery with remarkable accuracy. 
We find, in particular, that the inclusion of inter-site interactions in the corrective Hamiltonian
improves considerably the prediction of thermodynamic quantities when electronic localization
occurs in the presence of significant interatomic hybridization 
(as is the case for the Mn compound), and that the self-consistent evaluation of the effective interaction parameters
as material- and ground-state-dependent quantities 
allows the prediction of energy differences between different phases and concentrations. 
\end{abstract}

\maketitle

\section{Introduction}

The search for new and more performant materials for Li-ion batteries has received 
a strong impulse in the last decade, first due to the development and diffusion of
portable electronics and now also with a major focus on transportation and energy storage.
These applications impose multiple requirements 
on the materials: 
of being light-weight, environmentally friendly, of having high gravimetric/volumetric energy density, 
high power, fast rechargeability, long life, thermal/chemical stability, and low fabrication costs. 
Despite steady progress in recent years and the introduction
of new types of rechargeable batteries (e.g., K-, Na-, Mg-ion \cite{liu17,kubota18,kim17,guduru16}, or Li-air 
ones \cite{kojic12,jung16}) for dedicated purposes, many 
microscopic aspects of their behavior still need full clarification, 
with space for improvement and optimization \cite{yuan17,deng15,etacheri11,tarascon10,tarascon08};
significant research activity is taking place, in fact, on all battery components 
(anodes, cathodes, electrolytes) \cite{hayner12,nitta15,mekonnen16}.
The constant efforts to improve performance have stimulated a vigorous search for better 
materials, especially for electrolytes (in the attempt to design solid-state media able to 
sustain safely higher voltages than their liquid counterparts, and comparable ionic currents)
\cite{zheng18,meesala17,bachman16,ma16,lotsch17,goodenough15,kuhn14,goodenough17} and for 
cathodes (in order to identify more conductive, safer systems with higher energy density, higher 
voltage) \cite{tarascon10,frayret10,jow16,schipper17,lu16,lee17,twu15,lee14,kraytsberg12,li13,wang15,rozier15,goodenough13}.
Cathode materials are, in fact, particularly important for the improvement of rechargeable Li-ion 
batteries as these components are not only the source of power, 
but also embody some of the most critical bottlenecks
towards the improvement of current technologies 
including weight, safety, energy density, and overall power.

The relevance of the electronic and ionic degrees of freedom within the single particles or grains of the electrodes,
and their role in determining the performance of Li-ion batteries (e.g., rate capability, energy density), has made
the use of first-principles calculations fundamental in the understanding of their functionality
and increasingly more common for the characterization and design of battery 
materials \cite{twu15,hautier16,ceder98,ceder97,pigliapochi17,barbiellini12,maxisch06,ceder11,sgroi17,vanderven98,wolverton98,vanderven09_1,henkelman11,yiu13,tse11,bouchet12,huang12,morgan17,catti00,grey13,ceder04}. 
Since the open-circuit voltage corresponds to the redox potential of the electro-chemically
active species changing their oxidation state during the charge/discharge of the battery
(these are often transition-metal ions),
it is crucial for the energetics
of various phases and compositions involved in the charge/discharge processes
to be predicted accurately and reliably. 
High predictive accuracy is also needed
for other quantities besides the voltage and the stability 
of the different phases appearing at intermediate Li concentrations, 
such as the formation energies of defects,
or the viability of different doping strategies. 
A key difficulty in being quantitatively accurate in these predictions 
comes from the presence of transition-metal (TM) ions, typically present in variable oxidation states. 
In fact, most approximate exchange-correlation functionals used in current implementations 
of density-functional theory (DFT) \cite{hohenberg64,kohn65}, 
such as the local-density approximation (LDA) or the generalized-gradient
approximation (GGA), 
tend to over-stabilize delocalized states and 
are unable to capture accurately the localization of $d$ electrons
as a result of the remnant self-interaction errors present in functionals.
Often, these errors lead to a distinct failure in describing
the ground state of materials at intermediate Li concentrations, predicting
a metallic band structure and 
an even distribution of electronic charge on TM ions, rather than the
correct mixed-valence ground state, with 
electrons localized on a subgroup of TM ions at a lower oxidation state.
This overstabilization of the metallic state typically compromises the reliability of total energies
and the thermodynamics between different phases.
For these reasons DFT calculations on these materials require
functionals that 
are able to reduce or eliminate the spurious 
self-interaction that affects most of current approximations, 
to deliver a more pronounced localization on TM ions and
a faithful representation of mixed-valence ground states.
Unfortunately, these requirements are very difficult to satisfy 
for functionals of the electronic charge density alone; 
even the recently introduced SCAN semilocal meta-GGA density functional \cite{perdew_scan_15}, 
while very promising for various systems including layered oxides \cite{chakraborty18} 
and materials with well localized exchange-correlation holes 
\cite{ceder_mno2_scan_16}, 
is not fully satisfactory for general systems affected by the above-mentioned problems, especially
inter-metallic transition-metal compounds \cite{sun18,wolverton_scan_18}. 
The partial removal of the electronic self-interaction by 
adding a fraction of Fock exchange, as 
in hybrid functionals, greatly improves 
the localization of electrons \cite{urban16} (albeit at a significant computational cost), 
but might not solve problematic aspects related to energetics, as reported
in Ref. \cite{ceder11_1} and detailed below.

One meaningful alternative to correct for self-interaction errors is the DFT + Hubbard approach, where
the exchange-correlation functional is augmented by a Hubbard-model Hamiltonian acting on localized states
\cite{anisimov90,anisimov91,liechtenstein95, anisimov97,dudarev98,anisimov91_2,solovyev94,mazin97,pickett98}
(see in particular Refs \cite{kulik06,kulik08} for a discussion of self-interaction and DFT+U). 
This approach was first applied to the study of cathode materials by 
some of us (Refs. \cite{zhou04,zhou04_1,zhou04_2}); this work showed
that DFT+U with effective interactions computed from first-principles \cite{cococcioni05}, 
albeit averaged over different Li contents, can promote a more pronounced localization of $d$ electrons
and predict average voltages (with respect to Li/Li$^+$) in closer agreement
with available experimental data, while also recovering the correct thermodynamics
between phases. This effort led to Hubbard-corrected DFT becoming a standard
computational tool to perform predictive first-principles calculations on Li-ion cathode 
materials, and over the last decade its use on these systems has been broad and successful 
\cite{marianetti17,sgroi17,vegge16,sato16,dai14,bhowmik13,leoni12,yu11,maxisch06,zhou06,lebacq04}. 
Although the accuracy of the original application of DFT+U to these materials was largely due
to the possibility to compute the Hubbard parameter $U$ from first principles, its semiempirical evaluation
was often preferred, probably due to the complexity to evaluate $U$ reliably and efficiently
on every system of interest.
Recent computational studies on cathode materials conducted with various flavors
of DFT+U have also reiterated some key characteristics or shortcomings 
\cite{reboredo14,wolverton14}, including
$i)$ its dependence on its environment (structural, magnetic, and electronic phases); 
$ii)$ the impossibility to use material-dependent $U$'s in energy comparisons; 
$iii)$ the variation of $U$ in proximity of defects (e.g., impurities, local
deformations of the lattice, surfaces, etc); 
$iv)$ the difficulty to obtain a uniform improvement in the prediction of several properties with 
the same value of the Hubbard $U$ \cite{sato17};
$v)$ the scarce reliability of DFT+U in presence of significant 
hybridization between the metal cations and the neighbor anions. 
To circumvent these issues hybrid functionals have become an increasingly popular choice for
calculations \cite{urban16,wu13,nakayama16,johannes12}.
Although more computationally demanding, these functionals 
offer the advantage of containing at most one adjustable parameter, 
namely the fraction of exact (Fock) exchange, 
typically determined semiempirically \cite{urban15} and held constant for
all the systems analyzed.
However, the overall quality of results obtained when using hybrid functionals
turns out to be often comparable 
to those obtained with DFT+U or approximate DFT functionals \cite{urban16,ceder10}
with also some major qualitative failures, such as predicting a negative formation energy 
for Li$_{0.5}$FePO$_4$ \cite{ceder11_1}, as mentioned earlier. 
We also underscore that in the vision outlined in Refs. \cite{cococcioni05,kulik06} the point $i)$ above is
actually an intrinsic feature of DFT+U; in addition $U$ should always be considered 
pseudopotential-dependent, since it depends on the atomic Hubbard manifold on which it acts
and this is influenced very significantly e.g. by the oxidation state of the atomic 
all-electron reference calculation (see Appendix of Ref. \cite{kulik08}).

The present paper studies in detail a well known class of cathode materials 
- Li-metal phosphates with the olivine (orthorhombic) structure - 
with the aim of addressing some of the methodological issues outlined above. 
These compounds, whose chemical formula is 
Li$_{x}$MPO$_4$ (M = Fe, Mn in this work, with $x$ varying between 0 and 1 during the discharge 
of the battery), represent a particularly appealing family for Li-ion batteries cathodes \cite{padhi97}. 
In fact, the presence of stable
PO$_4$ tetrahedra as structural linkers improves the chemical stability of the cathode,
reducing the chance of oxygen releases, and guarantees higher levels of safety for the device, especially
if compared to that offered by other materials, such as LiCoO$_2$.
Among the olivine phosphates LiFePO$_4$ is of primary technological interest (in fact, it is
already largely used in the fabrication of commercial batteries) and is the object of intense research activity 
that aims at understanding its electronic properties and the surprisingly high (and still somewhat controversial)
charge/discharge rate that nanostructured cathodes are able to sustain 
\cite{grey14,niu14,maier14_1,bazant13,ogumi13,ceder13,huang12,yamada12,ceder11,bazant08}, 
despite its low ionic and electronic bulk conductivities. 
Ongoing research efforts are also targeting, at a more explorative level, 
other compounds of this family \cite{deng17}, with the main aim of improving the specific energy 
through higher voltages. While experiments assessing the performance of many of
these compounds as cathode materials are made difficult by the unavailability of electrolytes able to 
sustain voltages higher than 3.5 - 4 V, there still remain many aspects of their behavior to be 
clarified, especially related to the occurence and stability of other phases at intermediate Li content. Therefore,
first-principles calculations aiming at a precise assessment of the energetics, from which the average
voltage can be estimated \cite{ceder99}, and of possible intermediate phases, represent a 
particularly precious tool for the characterization of these materials and for the assessment of their performance.
As mentioned, this work will focus on two different olivine phosphates that have, respectively, Fe or Mn 
as transition-metal cations. In particular, it will aim at determining the equilibrium structure 
of these materials, the average
voltage with respect to the Li/Li$^+$ couple of a pure Li anode, the relative stability and the formation energy
of compounds with different Li contents ($x$ = 0, 0.5, 1). Total energy calculations will be performed
with an extended Hubbard-corrected functional (named DFT+U+V) that contains both on-site ($U$) and 
inter-site ($V$) effective electronic interactions \cite{campo10}. 
The work will clearly show 
the benefits of an improved 
description (thanks to the inter-site term $V$) of the hybridization between transition-metal ions 
and their oxygen ligands to capture electronic localization and mixed-valence ground states. 
The presented results will also highlight the critical importance to compute 
(specifically, from linear-response theory (LRT) calculations in supercells \cite{cococcioni05}), 
the Hubbard interaction parameters
in full consistency with the electronic ground state and the crystal structure of the materials considered
in order to achieve quantitatively reliable energetics.

The reminder of the paper is structured as follows.
Section II is devoted to describing the extended DFT+U+V functional and to 
discussing the self-consistent linear-response procedure adopted to compute the effective
electronic interactions, $U$ and $V$. 
Section III presents the results obtained for each of the 
materials considered, comparing them with available experimental data (e.g., voltages, equilibrium
crystal structures) and discussing the accuracy of different computational approaches.
Finally, some conclusions are proposed, highlighting the most important results and
the main merits of this extended Hubbard approach to study battery materials
and to perform quantitatively predictive total energy calculations.
In the Supplementary Information \cite{suppinf19} the results presented in the paper
are validated through a comparison with those obtained from slightly different Hubbard corrections
specifically, with finite Hubbard $U$ also on O $p$ states, Hubbard parameters evaluated
from linear-response calculations in larger supercells, or from a recent
implementation of LRT in density-functional perturbation theory (DFPT)
\cite{timrov18}.

\section{Methodology}

\subsection{The DFT+U+V functional and the self-consistent evaluation of $U$ and $V$ 
\label{upvc}}

The DFT+U+V method was introduced in Ref. \cite{campo10} as a generalization of DFT+U
(in its simplest, rotationally invariant formulation introduced by
Dudarev {\it et al.} \cite{dudarev98}), and it is based on an extended Hubbard model 
\cite{hubbardI,hubbardII,hubbardIII,hubbardIV,hubbardV,hubbardVI} that contains 
both on-site ($U$) and inter-site ($V$) electronic interactions. 
The total energy functional can be expressed as follows:
\barr
\label{uvfun}
E_{DFT+U+V} &=& E_{DFT} + \sum_{I,\sigma} \frac{U^I}{2}Tr\left[{\rm \bf n}^{II
\sigma}
\left({\mathbf 1}-{\rm \bf n}^{II\sigma}\right)\right]\nonumber \\
&-& \sum_{I,J,\sigma}^{\hspace{5mm}*}\frac{V^{IJ}}{2}
Tr\left[{\rm \bf n}^{IJ\sigma}{\rm \bf n}^{JI\sigma}\right]
\earr
where the atomic Hubbard manifold occupation matrices, obtained from the projection of Kohn-Sham (KS) states
on the atomic Hubbard manifold, have been generalized to allow for inter-atomic terms ($I\neq J$):
\beq
\label{occupij}
n^{IJ\sigma}_{m,m'}=\sum_{k,v}f_{kv}^{\sigma}\langle \psi_{kv}^{\sigma} |
\phi_{m'}^{J}
\rangle\langle \phi_{m}^{I} | \psi_{kv}^{\sigma}\rangle.
\eeq
In Eq. \ref{uvfun} $\sigma$ labels the spin of electrons while the star ``*" over the 
second summation is a reminder that, for each atom $I$, the sum runs only over neighboring atoms $J$ 
within a certain shell. The symbol ``$Tr$" indicates the trace of the matrix
to which it is applied, while the ${\mathbf 1}$ in the second term represents
the unitary matrix. 
In Eq. \ref{occupij} $f_{kv}^{\sigma}$ are the occupations of the KS states
$\psi_{kv}^{\sigma}$, labeled by a k-point index $k$ and by a band index $v$, while
$\phi_{m}^{I}$ are atomic states of the atom $I$, labeled by the magnetic quantum
number $m$. 
The present work only uses a finite inter-site interction $V$ between the $d$ states of
TM ions and the $p$ states of their oxygen neighbors.
Based on the above definitions, we note that Eq. \ref{uvfun} can also be rewritten more concisely as
\beq
\label{uvfun1}
E_{DFT+U+V} = E_{DFT} + \sum_{\sigma} {\bf Tr} \frac{\bf W}{2} \left[{\rm \bf n}^{\sigma}\left({\mathbf 1}-
{\rm \bf n}^{\sigma}\right)\right]
\eeq
where the ``${\bf Tr}$" operator is now understood to act also on implicit site indexes and the
interaction matrix ${\bf W}$ is such that $W_{II} = U^I$ and $W_{IJ} = V^{IJ}, \forall I\neq J$.

In order to understand how this extended corrective functional modifies the 
electronic structure of a system it is useful to study the action of the Hubbard
additional potential on a specific KS state. 
The generalized KS Hamiltonian can be obtained from the functional derivative of Eq. \ref{uvfun1}
with respect to the complex conjugate of the KS state:
\barr
\label{VUV}
&&\frac{\delta E_{DFT+U+V}}{\delta (\psi^{\sigma}_{kv})^*}= 
V_{DFT+U+V}|\psi_{kv}^{\sigma}\rangle  =
V_{DFT}|\psi_{kv}^{\sigma}\rangle \nonumber \\
&&+ \sum_I \frac{U^I}{2}\sum_{m,m'}\left(\delta_{mm'} -
2n^{II\sigma}_{m'm}\right)|\phi^{I}_{m}\rangle\langle\phi^{I}_{m'}|\psi_{kv}^{\sigma}\rangle \nonumber \\
&&- \sum_{I,J}^{\hspace{5mm}*} V^{IJ}
\sum_{m,m'}n^{JI\sigma}_{m'm}|\phi^{I}_{m}\rangle\langle\phi^{J}_{m'}|\psi_{kv}^{\sigma}\rangle.
\earr
From the equation above it can be seen that while the on-site part of the potential discourages partial atomic occupations, 
stabilizing states that are either full or empty, 
the inter-site term favors states having significant overlap with atomic states
of neighboring atoms (thus contributing substantially to $n^{JI\sigma}_{m'm}$ with
$I\neq J$).
Given this competition between the two types of interactions 
an accurate and consistent evaluation of both parameters is crucial.
In our work this task is achieved by using the linear-response approach for the calculation of
$U$, introduced in Ref. \cite{cococcioni05}, and extending it to the calculation 
of $V$, as discussed in Ref. \cite{himmetoglu14}.
Within this approach the inter-site interaction $V$ can be obtained as the
off-diagonal element of the interaction matrix of which $U$ represents the 
diagonal part; this has already been shown to be accurate 
in a number of different cases,
from ionic oxides and covalent semiconductors \cite{campo10} to transition-metal complexes \cite{kulik11}. 

In order to fully capture the dependence of the effective interactions on the 
electronic and crystal structures, in the present work we adopt a self-consistent
procedure for their calculation that is an evolution of the one introduced in Ref. \cite{kulik06}
and, already discussed in Refs \cite{campo10} and \cite{himmetoglu14},
has been used in a number of other studies afterwards \cite{hsu11,yu12,shukla15,shukla16,mann16}.
A similar self-consistent approach was introduced, independently, in Ref. \cite{sato16},
albeit with some differences that will be highlighted below.
The rationale of this approach relies on the evaluation of $U$ and $V$ using linear-response theory 
from a DFT+U+V ground state until self-consistency.
In practice, starting from an initial choice of these parameters 
$U_{in}^1$ and $V_{in}^1$ (possibly equal to 0) a sequence of 
linear-response calculations is started in which the interactions obtained from the
$i^{th}$ step are used to generate the DFT+U+V ground state of the next one: 
$U_{in}^{i+1} = U_{out}^{i}$, $V_{in}^{i+1} = V_{out}^{i}$ (mixing schemes between
``input" and ``output" can be adopted to improve convergence).
The sequence is terminated when input and output values of the effective
interactions coincide within a numerical threshold.
In this work we also 
perform a structural optimization (of both cell parameters and atomic positions) 
in between two successive calculations of the interactions parameters, in order to guarantee
full consistency of their values also with the crystal structure. Indicating the
interaction parameters with the general symbol $W$ ($W = \{U,V\}$), the self-consistent procedure can 
be summarized as follows: 
\begin{widetext}
\begin{center}
$W_{in}^1$ $\rightarrow$ SO($W_{in}^1$) 
$\rightarrow$ LR($W_{in}^1$) $\rightarrow$
$W_{out}^1 = W_{in}^2$ $\rightarrow$ SO($W_{in}^2$) 
$\rightarrow$ LR($W_{in}^2$) $\rightarrow$ $W_{out}^2 = W_{in}^3$ $\ldots$
\vspace{4mm}
$\ldots$ $W_{out}^i = W_{in}^{i+1} = W_{in}^{i}$ and ES($W_{in}^{i+1}$) = ES($W_{in}^{i}$) 
\label{uvscf}
\end{center}
\end{widetext}
where ``SO" stands for structural optimization and ``ES" for equilibrium (crystal) 
structure. The sequence is terminated when also the equilibrium
crystal structure is converged. This procedure, which evolves the effective
interactions with the ground state of the system, allows to
treat them as material- and environment-specific quantities (rather than as parameters 
of the calculation, as in most literature on Hubbard corrections, often oblivious to
the fact that Hubbard parameters need to be pseudopotential-specific)
and this is one of the key aspects that will be discussed in this work.

It is important to stress that the approach chosen for the self-consistent calculation
of $U$'s and $V$'s is slightly different than that
adopted in Ref. \cite{sato16}: in fact, in the LRT calculations starting from a DFT+U+V ground state,
the potential deriving from the Hubbard functional is kept fixed 
(computing it on unperturbed atomic occupation matrices) to make sure that
only the DFT part contributes to the second derivative that defines
the effective interactions. This procedure makes the evaluation of $U$ and $V$
consistent with their definition as effective spurious curvatures of the DFT energy
with respect to atomic occupations \cite{cococcioni05}. It is also consistent with
the use of newly obtained values as a new guess for the next iteration,
rather than as a correction to those of the previous step.
A thourough comparative analysis of various self-consistent procedures for computing 
the Hubbard interaction parameters has been recently proposed in Ref \cite{oregan17}. It is useful to
remark that, since the Hubbard potential during the calculation of the Hubbard
parameters is fixed, the present work (as well as that of Ref. \cite{campo10})
is consistent with and gives the same result of method 2 of Ref. \cite{oregan17} 
(the $U_{scf}$ mentioned here corresponds to $U^{(2)}$ of that work).
In fact, it can be proven that the 
Hubbard parameters calculated through the procedure just illustrated corresponds to the (atomically averaged)
matrix element of the Hartree and exchange-correlation kernel, computed at the DFT+U ground
state with $U = U_{in}$.

For the sake of a precise numerical comparison of the Hubbard interaction parameters obtained it
would be appropriate to discuss screening of these quantities and how it is accounted for
during their evaluation. This discussion is sketched in Ref. \cite{oregan18} presenting a linear-response
calculation of the Hubbard parameters equivalent to (and based on) the one introduced in Ref. \cite{cococcioni05}.
We remark here that the perturbation is screened by all the ``other" electronic states, i.e. those that are not 
explicitly perturbed by a shift in the potential acting on them.
This implies that, in principle, the Hubbard $U$ calculated for DFT+U and DFT+U+V
ground states can be expected to be different from one another because the necessity to perturb
neighbor ligand states for the evaluation of the inter-site $V$ leads, for the latter, to the
removal of these states from the ``screening" manifold.
For most calculations presented in this work we decided
to adopt a uniform strategy for computing the
Hubbard interaction parameters and to always include the response of relevant ligand states
(typically, O 2p). The validity of this choice was then assessed for DFT+U calculations
on Li$_x$FePO$_4$ by comparing the values of $U$ obtained as outlined above
with those from calculations that did not include perturbations of O atoms.
A further confirmation of the quality of this approximation is provided
by the comparison with calculations of the Hubbard parameters based on DFPT\cite{timrov18}, 
as is detailed in the Supplementary Information \cite{suppinf19}.

\subsection{Formation energies and voltages}

Most of the results presented in this work concern the evaluation of 
formation energies and (average) voltages, of which this section provides a definition.

Given a generic system $S$ 
(able to reversibly intercalate Li ions in its structure)
the formation energy of the compound Li$_x$S ($0 \leq x \leq 1$) with respect
to the LiS and S component is evaluated as a weighted difference of
total energies: 
\beq
E_f(x) = E(Li_xS) - x E(LiS) - (1-x) E(S).
\label{fe}
\eeq
This formula is approximate at finite temperatures, where
the difference between free energies (rather than total/internal energies)
should be considered. 
While widely used in literature, the cancellation between the entropic terms seems a 
reasonable assumption for the vibrational part, while the configurational 
(e.g., ionic, electronic, magnetic) contributions to the entropy should, in principle, be handled
with care.
This is a particularly difficult task for intermediate Li concentrations. In fact, while
for fully lithiated and delithiated compounds ($x = 0$ and $x=1$) this only involves 
screening different spin configurations of the transition-metal ions, at fractional
values of $x$ additional important terms arise from the Li fractional occupation of available
sites \cite{ceder00,zunger87,marzari94,asta01} and from the electronic ``configurational entropy" related to the localization of 
valence $d$ electrons on a subset of transition-metal ions. This latter term
can in fact determine the stabilization of intermediate compounds, as previously 
discussed for Li$_x$FePO$_4$ \cite{zhou06}.

In this work, following most of the literature on the topic, the evaluation of the stability
of compounds at intermediate Li content will be based on total energies. A significant effort
has been focused however on capturing the localization of electrons on transition-metal ions
(i.e. the differentiation of these ions in subgroups of different valence) and on the 
quantitative comparison of different Li configurations at $x = 0.5$.

Another important quantity that can be evaluated from the direct comparison of total energies
is the voltage (i.e. the Li intercalation potential) with respect to an 
ideal Li/Li$^{+}$ anode.
For a generic system $S$, considered as a cathode material, this quantity is in general
a function of the Li content (although most valuable systems show no/very low dependence of 
the voltage on $x$, at least for a good part of the admissible range). 
Its average value, between Li concentrations 
$x_1$ and $x_2$ ($x_2 > x_1$) can be computed as:
\beq
\phi_{x_1,x_2} = -\frac{E(Li_{x_2}S) - E(Li_{x_1}S) - (x_2 - x_1)E(Li)}{(x_2 - x_1)e}.
\label{volt}
\eeq
The total energies entering this formula are obviously referred to the same amount of 
material (e.g., one formula unit or one unit cell of the crystal) and $E(Li)$ represents
the total energy of an equivalent number of Li atoms in bulk Li (representing the
anode), while $e$ stands for the electronic charge. 
As noted for the calculation of formation energies and also for voltages, 
total energies should be replaced by total free energies; however, it is a common practice to compare voltages
obtained from total energies with the results of experiments at finite T. 
The assumption of cancellation of the entropic terms in the difference between
free energies is on firmer ground than in the evaluation of formation energies
if voltages are computed across the whole range of variation of Li content, i.e. 
between $x_1 = 0$ and $x_2 = 1$,
because then no calculation is needed at intermediate Li concentrations: 
for both the fully lithiated and delithiated compounds
the configurational entropy due to either electrons or Li ions vanishes, while
terms related to spin configurations, broadly independent from the Li content,
presumably cancel.
In this work we will adopt this common practice and will evaluate the average voltages
balancing the total energies of fully lithiated and delithiated compounds.
This quantity assumes a particularly simple expression that corresponds to
the formation energy of the fully lithiated compounds (with respect to the delithiated
one and bulk Li) normalized by the electronic charge:
\beq
\phi = -\frac{E(LiS) - E(S) - E(Li)}{e}.
\label{volt}
\eeq

\section{Technical details} 
\label{tecdet}
The first-principles calculations presented in this work are all performed using the
pseudopotential, plane-wave implementation of DFT contained in the 
Quantum ESPRESSO distribution \cite{qe09,qe17}. In all cases a generalized-gradient approximation 
(GGA) functional is chosen, constructed with the    
PBEsol parametrization \cite{perdew_sol}, particularly well suited for crystalline solids.
In the calculations discussed in this work different systems were modeled
with pseudopotentials of different types:
the Fe olivine was described with ultrasoft pseudopotentials \cite{Vanderbilt1990}
taken from the Pslibrary 1.0.0 \cite{pslib1.0} \footnote{The following pseudopotentials 
were used: Fe.pbesol-spn-rrkjus-psl.0.2.1.UPF, 
O.pbesol-n-rrkjus-psl.0.1.UPF, P.pbesol-n-rrkjus-psl.0.1.UPF,
Li.pbesol-s-rrkjus-psl.0.2.1.UPF}, while for the 
Mn compound PAW pseudopotentials \cite{blochl94} were taken
from the Pslibrary 0.3.1
\cite{pawlib14} \footnote{The following pseudopotentials 
were used: Mn.pbesol-spn-kjpaw-psl.0.3.1.UPF,
P.pbesol-n-kjpaw-psl.0.1.UPF, O.pbesol-n-kjpaw-psl.0.1.UPF, Li.pbesol-s-kjpaw-psl.0.2.1.UPF}
following the Standard Solid State Pseudopotentials library (SSSP) validation protocols \cite{sssplib18}. 
More detailed informations about the pseudopotentials and the parameters used in the
calculations are given in the sections dedicated to presenting results on specific
systems. 

\begin{figure}[h!]
\begin{center}
\subfigure[]{
\includegraphics[width=8cm]{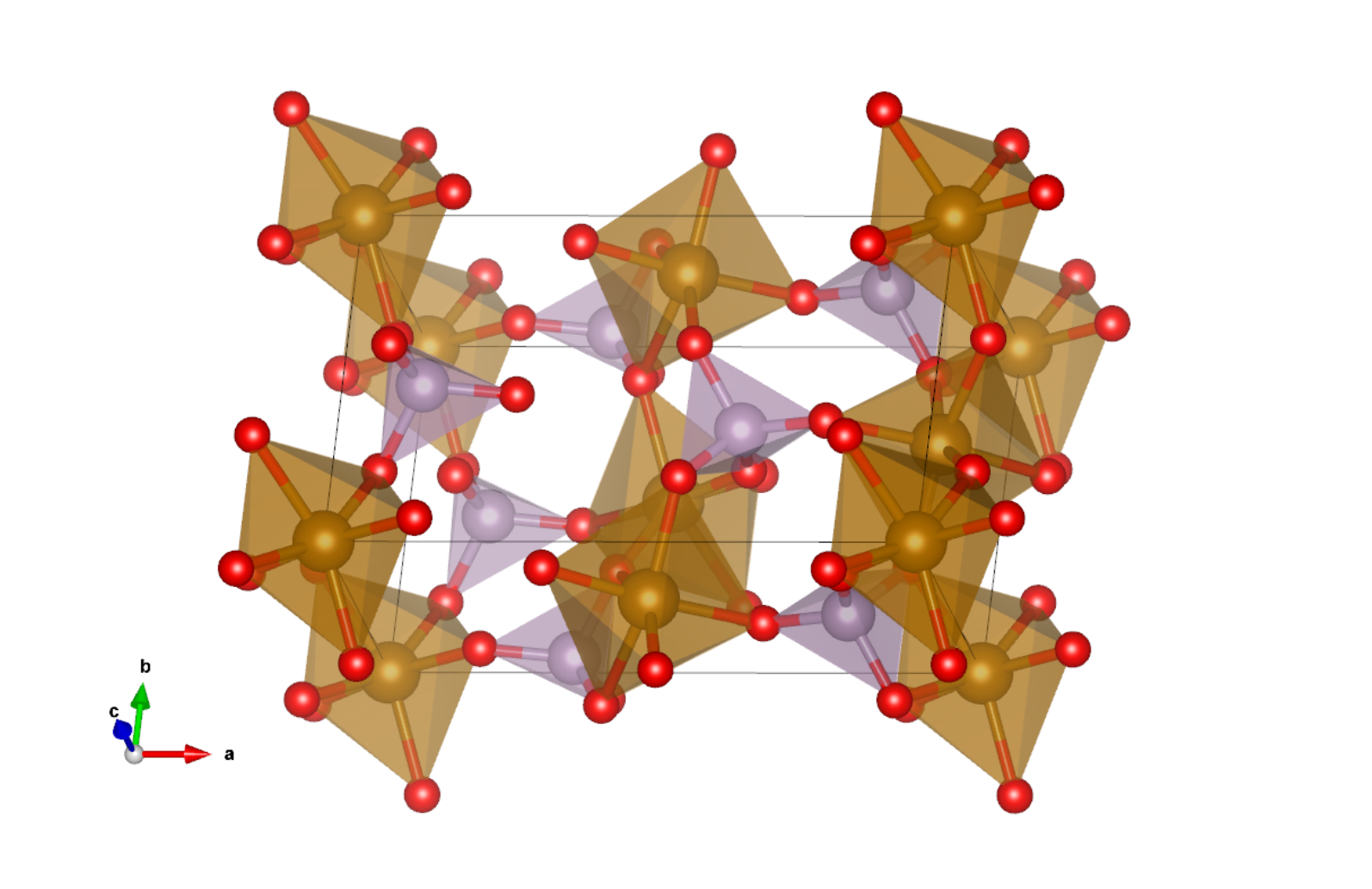}}
\subfigure[]{
\includegraphics[width=8cm]{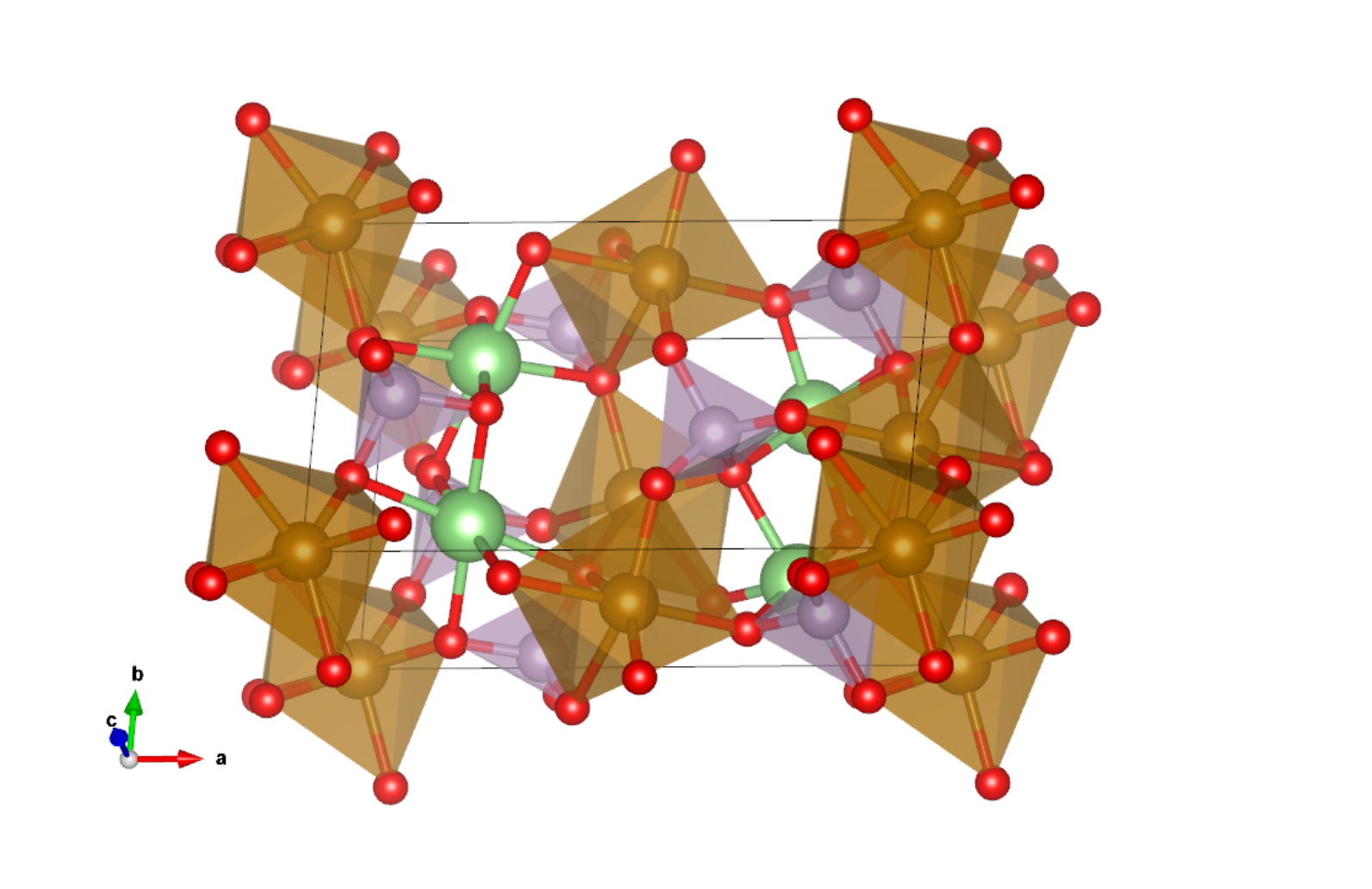}}
\caption{\label{lfpostruc} The unit cell of FePO$_4$ (a) and LiFePO$_4$ (b). In 
evidence both the corner-sharing Fe-O octahedra and the P-O tetrahedra acting as linkers between them. 
The comparison between the two cells highlights the position of Li ions along channels parallel to 
the $b$ axis of the orthorhombic cell.} 
\end{center}
\end{figure}

The orthorhombic unit cell of the olivine materials considered in this work
contains four formula units (24 to 28 atoms, depending on Li content), 
and is large enough to accommodate few antiferromagnetic configurations 
(detailed and compared for selected compositions, see the following
material-specific sections) and five Li concentrations (specifically, $x=0, 0.25, 0.5, 0.75, 1$) of
which $x=0, 0.5, 1$ will be explicitly considered in this work.
As an illustrative example,
Fig. \ref{lfpostruc} compares the $x=0$ and $x=1$ crystal structures (unit cells) of the Fe system.
The unit cell of transition-metal phosphates is often attributed the $Pnma$ space group with 
the longest dimension along the $x$ axis, 
and the $y$ direction along the chains Li ions form when fully occupying the 
available sites.
As can be observed from Fig. \ref{lfpostruc}, transition-metal ions are coordinated by six oxygen 
octahedra that are distorted from their ideal, cubic symmetry and tilted with respect to
neighbor ones. TM-O octahedra are connected to each other either by sharing apical oxygens or by PO$_4$ 
tetrahedra which also share apical oxygens and act as structural linkers.
This structural framework, which is the same also in the Mn phosphate, 
undergoes relatively minor but TM-dependent distortions 
upon Li insertion. This structural stability (together with
the chemical one) is one of the main aspects that makes these crystals good cathode materials, as
it enables the possibility to undergo several lithiation cycles with minimal structural
damage.
Given the proportions of the unit cell (roughly twice as long along $a$ than along $b$ and $c$),
integrations over the Brillouin zone of this lattice
were determined to require a 2$\times$4$\times$4 a grid of special k-points \cite{monkhorst76}
along with a Methfessel-Paxton \cite{methfessel89} or gaussian 
smearing of the Fermi occupation function of 10$^{-2}$ Ry.
Since the system has an insulating ground state, in principles it could be treated within a fixed-occupations
scheme. 
However, starting the self-consistent calculation of the Hubbard
parameters from a DFT ground state, typically metallic, requires the use of a finite smearing of 
occupations that also facilitates, from a technical point of view, the description of magnetic systems.

The same four-formula-units supercell is also used 
to compute the Hubbard interaction parameters using the linear-response method
introduced in \cite{cococcioni05}. For DFT+U+V calculations the effective interactions 
of the Fe and Mn olivines are also computed in a 1$\times$2$\times$2 supercell of the one 
described above, to study the convergence of the Hubbard interactions with the cell size. 
It is important to remark that, as already discussed in Ref \cite{campo10}, computing $V$ does not
require any modification to the linear-response (LR) approach introduced in \cite{cococcioni05} 
(in fact, $V$ is obtained as the off-diagonal element of the atomic interaction
matrix); however, it implies perturbing also non transition-metal states,
which increases proportionally the cost of the LR calculations.
In the present work, since calculations include inter-site interactions between
transition-metal ions and their first neighbors, oxygen atoms are also individually
perturbed. More specifically, perturbations are imposed on the 3$d$ states of 
transition-metal ions and on the 2$p$ states of oxygens. 
While perturbing oxygens' $p$ states is strictly necessary only for DFT+U+V calculations, the procedure
outlined above was adopted also for on-site-only DFT+U calculations.
A second series of calculation of the Hubbard parameters was performed for both 
the DFT+U and DFT+U+V cases, using a
new implementation of the linear-response method based on density-functional 
perturbation theory (DFPT) \cite{timrov18}. 
As detailed in the Supplementary Information \cite{suppinf19}
this DFPT-based approach overall confirms (typically adding a slight quantitative refinement)
the results discussed in the next sections, obtained from the smallest (4-formula units) cell,
even if deviations of the computed values of the Hubbard parameters are occasionally 
observed.

As mentioned in the previous section, the interaction parameters of the corrective
Hubbard functional are computed self-consistently with respect to both the electronic
and crystal structures. It is worth keeping in mind that while DFT+U+V total energy
calculations (including the ones linear-response calculations are based on) are performed
on an orthogonalized Hubbard basis, structural optimizations are performed 
without orthogonalizing it. This inconsistency, due to the technical difficulty
of implementing forces and stresses from DFT+U+V in an orthogonal atomic
basis set, is expected to lead to marginal errors, and the crystal structures 
computed in this paper can be considered essentially identical to the full self-consistent ones.
In any case this is common practice in all implementations of DFT+U. 

\section{Results}

\subsection{Li$_x$FePO$_4$}
\subsubsection{Electronic structure and energetics}
\label{fe}
Among TM phosphates with the olivine (orthorhombic) structure, Li$_x$FePO$_4$ is currently the most widely
used in commercial Li-ion batteries and certainly one of the most
intensely studied in literature. In fact, after it was proposed as a potential cathode material
in 1997 \cite{padhi97}, due to its relatively high capacity, superior thermal, chemical and mechanical 
stability, low cost and environmental-friendliness of its components, this material has become the 
object of a fervent research activity
aimed at improving the understanding of its properties and its behavior when employed in actual devices.
While the use of nanostructured cathodes
has allowed to overcome the low ionic and electronic conductivities of this material
(thus permitting its large-scale deployment in commercial devices), the 
microscopic diffusion mechanisms
\cite{barbiellini12,craco11,leoni11,maxisch06,mccarty13,henkelman11,tse11,
masquelier08,yamada12,weill08,nazar06,maier08}
and the nucleation, relative stability and transformation dynamics of various phases
\cite{zhou06,laffont06,huang12,maier09,maier12,maier11,maier14_1,maier15,grey14,yiu13,niu14,ogumi13,masquelier05,bazant12}
occurring during the charge and discharge transients of the battery
are still matter of intense research. 
In fact, in light of the very limited miscibility of Li-rich and Li-poor phases (somewhat 
increased at the nanoscale due to the higher free energy costs for
the creation of interfaces \cite{vanderven09,vanderven09_1,heinonen15}), 
the explanations of the relatively fast charge and discharge cycles observed in nanostructured 
cathodes is still quite controversial, as it is expected to involve out-of-equilibrium
phenomena such as phase-transformation wave dynamics \cite{bazant08}, metastable  
intermediate phases \cite{maier09,maier11,maier14_1,grey14}, 
or solid solutions \cite{ceder11,niu14,ceder13}.
This difficulty in interpreting the observed behavior constitutes a further 
motivation to precisely assess the energetics,
the equilibrium crystal structure and the relative stability of the material at
different Li concentrations, which is the main scope of this work.

As mentioned in Section \ref{tecdet} 
the transition-metal phosphates studied here 
are modeled with an orthorhombic 24-atoms unit cell (unless explicitly indicated)
that corresponds to four formula units (see Fig. \ref{lfpostruc}).
This cell is large enough to accommodate a few antiferromagnetic states and Li configurations
which will be compared in the last part of this section.
While the energy scale of inter-atomic magnetic interactions makes the comparison
of various AFM configurations marginally relevant
for the energetics of the material (in particular, for its average voltage and the
formation energy of the half-lithiated compound) especially at finite temperatures,
this comparative analysis is still interesting to better understand the electronic structure
of cathode materials \cite{jena16} and, in general, to improve their characterization
(in comparison with experiments) \cite{singh15,julien06}, especially if a significant
shift of magnetic properties (e.g., the N\'eel temperature) is expected to correlate 
with Li intercalation. 
All the calculations shown in the first part are performed for
the antiferromagnetic ordering (named AF1 in the later discussion)
that was verified to correspond to the ground state configuration.
The comparative evaluation of different phases of the partially lithiated system is, instead, 
more important to assess cathode performance.
In fact, the mechanisms and kinetics of the (de-)lithiation process during the charge and 
discharge transients are governed by how Li ions intercalate into the structure 
of the partially lithiated material. In spite of very limited miscibility, 
half-lithiated phases have been reported
to form at the interface of the $x = 1$ and $x = 0$ regions \cite{maier11,maier14_1,grey14}, 
probably stabilized by the partial relaxation
of the misfit stress that they allow, mediating the lattice mismatch between the two end phases
\cite{vanderven09,vanderven09_1}.
While the topic is still controversial (the thickness and orientation of interfacial half-lithiated phases 
is observed to depend on the size of the particles \cite{maier11,maier14_1}
and possibly on the operational conditions 
\cite{ceder13}),
assessing the relative stability (i.e. the energy of formation) and comparing the crystal structures 
of crystals with intermediate Li content with those of the fully lithiated and de-lithiated
compounds provides relevant information to evaluate the energy cost to form these
interfaces, the magnitude of strains in their surrounding that might contribute to their
stabilization, and the kinetics of Li ions inside the electrodes' bulk. 
In this work we will consider two specific half-lithiated crystals, both alternating fully occupied
and unoccupied Li planes, that are parallel to the $yz$ plane in the first case, and to the $xz$ plane in the second one.
These structures will be called ``$yz$" and ``$xz$", respectively.
Since a full set of calculations with all the various flavors of Hubbard corrections was performed only for
the first of the half-lithiated structures, the second will be considered only in the last part of the 
section, in the context of a quantitative comparison between the two. 

The calculations on the Fe olivine were performed using an ultrasoft pseudopotential for Fe 
(with semicore 3$s$ and 3$p$ states in valence, along with 4$s$ and 4$p$) that required 
kinetic energy cut-offs of 85 and 650 Ry for the electronic wavefunctions and charge density, respectively.
In this section we compare the results obtained within various approximations, all based
on the same GGA (PBEsol) functional, which are labeled as GGA, DFT+U$_{ave}$, DFT+U,
DFT+U+V. 
For the sake of completeness, in section \ref{SI.slfpo} of the Supplementary Informations \cite{suppinf19} we 
also present results 
obtained from DFT+U+V with a finite $U$ on the $p$ states of oxygen, with the
Hubbard parameters computed from a larger 1$\times$2$\times$2 supercell, or from DFPT
(named, respectively, DFT+U$^{dp}$+V, DFT+U+V$_{1 \times 2 \times 2}$, DFT+U$_{DFPT}$ and DFT+U+V$_{DFPT}$).
DFT+U$_{ave}$ and 
DFT+U are both based on standard Hubbard corrective functionals (i.e. the first two 
terms in Eq. \ref{uvfun}); however, the Hubbard U is computed in two different ways, one common in the 
literature, and one presented here. 
In DFT+U$_{ave}$, GGA is used to optimize the structure
at $x = 0$, 0.5 and 1. The Hubbard $U$ is then computed from a single-shot LR calculation for each of 
these three crystals in their respective equilibrum configurations and averaged over them. 
The average value of the Hubbard U obtained in this way is then used to calculate the total energy
of the crystal for the three Li concentrations and
to evaluate the (average) voltage and the formation energy of the $x=0.5$ compounds,
without further optimization of the crystal structure. This is the common procedure
followed in literature when the Hubbard $U$ is computed from first principles
rather than being determined semi-empirically by fitting existing experimental
data; for this reason the DFT+U$_{ave}$ results will be used as a benchmark
with respect to those from the other corrective functionals used in this work. Instead, in DFT+U (and also
in DFT+U+V) 
the Hubbard interaction parameters are computed 
self-consistently (both with the electronic and crystal structures), according to the procedure
described previously in section \ref{upvc}.
In particular, the inter-site Hubbard interactions ($V$) are computed 
and used between Fe and nearest-neighbor oxygen ions. 
In these cases, no averaging is performed 
and all energy balances are obtained from calculations using material-specific 
and Li-concentration dependent $U$'s and $V$'s.
%
\begin{widetext}
\begin{center}
\begin{table}[h!]
\begin{tabular}{|c|c|c|c|c|}
\hline
\hline
 & Interaction & LiFePO$_4$ & \multicolumn{1}{c|}{Li$_{0.5}$FePO$_4$} & \multicolumn{1}{c|}{
FePO$_4$}  \\
\hline
\multirow{1}{*}{DFT+U$_{ave}$} & U$_{Fe}$ & \multicolumn{3}{c|}{6.93} \\
\hline
\multirow{1}{*}{DFT+U} & U$_{Fe}$ & 4.88 & 5.08 / 5.53 & 5.21\\
\hline
\multirow{2}{*}{DFT+U+V} & U$_{Fe}$ & 5.14 & 5.44 / 5.05 & 5.30\\
& V$_{Fe-O}$ & 0.30 - 0.88 & 0.28 - 1.12 / 0.39 - 0.82 & 0.52 - 1.12 \\
\hline
\hline
\end{tabular}
\caption{The values of $U$s and $V$s (in eV) for the three Li concentrations considered, 
computed within various flavors of Hubbard-corrected functionals.
The ranges of values reported for the $V$ parameters refer to different O ions in the first
coordination shell, since values vary with the M-O distance.
For Li$_{0.5}$FePO$_4$ the two sets of values refer to the Fe$^{2+}$ and Fe$^{3+}$ ions,
respectively.}
\label{uev}
\end{table}
\end{center}
\end{widetext}
Table \ref{uev} shows the values of all the interaction parameters computed
for each of the Li concentrations considered for the Fe olivine and
for the flavors of the Hubbard corrective functionals outlined above.
Additional results can be found in Table \ref{SI.uev}.
It is easy to note that the value of $U$ obtained from a single-shot LR calculation
on the GGA ground state (and averaged) is significantly higher than all the others,
which is probably a consequence of the lack of consistency with the 
electronic ground state.
Substantial differences between the
self-consistent values of the Hubbard parameters can also be noted
comparing the results at different Li concentrations (i.e., in dependence of the level of
oxidation of the metal ion) or the results of different approaches when
the same material or the same oxidation state of Fe are considered. 
For example, the value of $V$ between Fe$^{2+}$ 3$d$ and O 2$p$ states changes 
depending on whether the Fe ion is one of LiFePO$_4$ or
of Li$_{0.5}$FePO$_4$. 
Equally significant differences can in fact be noted comparing the values of
$U$ obtained for Li$_{0.5}$FePO$_4$ and FePO$_4$ within different flavours of the 
Hubbard correction or for Fe ions in the $x=0.5$
compound and those in the same oxidation state in either fully lithiated
or delithiated materials.  
These differences, which
are the result of the self-consistent procedure adopted in the calculation
of $U$'s and $V$'s, confirm a sensitivity of their values on the fine 
details of the chemical and crystal environments of the transition-metal ions 
they refer to, and suggest a limited portability of these parameters from one
system to another, even when treated within the same approximations. 

Based on these observations, we argue that the Hubbard interactions 
should not be considered as parameters of the calculation, nor should be thought of
as simple functions of average quantities.
We prefer instead to view these as 
quantities that depend on the electronic structure of the system,
whose value is determined self-consistently by the ground state they contribute to determine. 
This point of view justifies our self-consistent calculation of the
Hubbard interaction parameters, which is further supported by the overall 
agreement of the equilibrium crystal structure (Table \ref{lattice_lfpo}) with the
experimental data for DFT+U+V and the significant
improvement over DFT+U obtained with the introduction of the inter-site interaction
$V$ in the energetics and electronic structure (see later), in spite of the quite broad 
range of self-consistent values for $U$'s and $V$'s.
\begin{widetext}
\begin{center}
\begin{table}[h!]
\begin{tabular}{|c|c|c|c|c|c|}
\hline
\hline
 & & \multicolumn{1}{|c|}{GGA} & \multicolumn{1}{c|}{DFT+U} & \multicolumn{1}{c|}{DFT+U+V} & Exp \\ 
\hline
\multirow{3}{*}{LiFePO$_4$} 
& a & 19.31 & 19.57 & 19.52 & 19.54$^a$/19.53$^b$ \\
& b/a & 0.58 & 0.58 & 0.58 & 0.58$^{a,b}$\\
& c/a & 0.46 & 0.45 & 0.45 & 0.45$^{a,b}$ \\
\hline
\multirow{3}{*}{Li$_{0.5}$FePO$_4$} &
a & 18.87 & 18.92 & 19.08 &  \\
& b/a & 0.59 & 0.59 & 0.59 & \\
& c/a & 0.48 & 0.48 & 0.47 &  \\
\hline
\multirow{3}{*}{FePO$_4$} & 
a & 18.64 & 18.69 & 18.61 & 18.44$^a$/18.56$^b$ \\
& b/a & 0.59 & 0.59 & 0.59 & 0.59$^{a,b}$ \\
& c/a & 0.49 & 0.49 & 0.49 & 0.49$^{a,b}$ \\
\hline
\end{tabular}
\caption{The equilibrium lattice parameters (in bohr) of Li$_x$FePO$_4$, $x = 0$, 0.5, 1,
computed with DFT (GGA at the PBEsol level) and with the Hubbard +U and +U+V corrections,
and compared with available experimental values (the superscripts $a$ and $b$ indicate Ref. \cite{rousse03} and Ref. 
\cite{padhi97}, respectively).}
\label{lattice_lfpo}
\end{table}
\end{center}
\end{widetext}
As apparent from Table \ref{lattice_lfpo},
the DFT+U functional tends to expand the equilibrium lattice parameters with 
respect to those obtained from GGA calculations. The presence of inter-site interactions partially
counteracts this tendency, and mitigates the effect of the on-site 
effective repulsion $U$. 
For LiFePO$_4$, compared to available experimental data of Ref. \cite{rousse03} and \cite{padhi97}, 
GGA uncharacteristically underestimates the equilibrium lattice parameters while DFT+U produces
an optimized unit cell in quite good agreement with the data.
This result is further refined by DFT+U+V.
Because of the experimental difficulties in stabilizing any compound at intermediate Li concentration, 
no measurement of the equilibrium lattice parameter of 
Li$_{0.5}$FePO$_4$ is, to the best of our knowledge, available. 
As for the $x=1$ case, we can observe that DFT+U leads to an expansion of the
equilibrium lattice parameters compared to GGA. However, at variance
with the fully lithiated case, the addition of the inter-site interaction $V$
yields a further expansion of the lattice.
It is useful to mention that the unit cell of the $x=0.5$ compound 
considered here undergoes a monoclinic 
distortion and the angle $\beta$ between $a$ and $c$ decreases its amplitude from
90$^{\circ}$ to about 88$^{\circ}$ (the variations between different approaches
are negligible).
For FePO$_4$, GGA equilibrium lattice parameters slightly overestimate
(by less than 1\%, in most cases) the experimental values. DFT+U further expands lattice
parameters with largest effects on the 
(longest) $a$ axis.
The addition of the inter-site coupling generally improves
the prediction of the cell parameters over the GGA results
and produces the best agreement with available experimental data.
\begin{widetext}
\begin{center}
\begin{table}[h!]
\begin{tabular}{|c|c|c|c|c|c|}
\hline
\hline
  & Fe$^{2+}$ ($x = 1$) & Fe$^{2+}$/Fe$^{3+}$ ($x=0.5$) & Fe$^{3+}$ ($x = 0$) & F. E. (meV/f.u.) & Voltage (V) \\
\hline
GGA & 6.33 & 6.11/6.08 & 5.93 & -126 & 2.72 \\
\hline
DFT+U$_{ave}$ & 6.18 & 6.19/5.68 & 5.65 & 161 & 4.09 \\
\hline
DFT+U & 6.20 & 5.74/6.19 & 5.71 & 191 & 3.83 \\
\hline
DFT+U+V & 6.22 & 6.22/5.77 & 5.76  & 107 & 3.51 \\
\hline
Exp & & & & \textgreater~0 & $\sim 3.5$ \\
\hline
\end{tabular}
\caption{L\"owdin total occupations of Fe 3$d$ states, formation energy, and average voltage 
computed with different methods for Li$_x$FePO$_4$, in comparison with available experimental data.
It should be stressed how the mixed-valence occupations are described very accurately, in addition to the voltage.}
\label{fev}
\end{table}
\end{center}
\end{widetext}
The key results of this work for Li$_x$FePO$_4$ are displayed in
table \ref{fev}, that shows 
the total occupation of the $d$ states for Fe ions (the trace of the diagonal
blocks of the matrix defined in Eq. \ref{occupij}), the formation energy
of Li$_{0.5}$FePO$_4$, and the average voltage with respect to Li/Li$^+$, computed as
indicated in Eq. \ref{volt}.
Analyzing the atomic occupations it is evident how the GGA (PBEsol) functional fails
to capture the charge disproportionation that should be observed upon delithiation
and the consequent differentiation of Fe ions into 2+ and 3+. In fact, one should observe the occupations in
the mixed-valence Fe$^{2+}$/Fe$^{3+}$ case mirroring those of the pure 2+ or 3+ cases. 
This failure results from the incomplete localization of valence electrons on the
$d$ states due to the defective cancellation of the electronic self-interaction,
that is a well known problem of most approximate DFT functionals.
DFT+U is effective in fixing this problem 
\cite{zhou04,zhou04_1,zhou04_2,barbiellini12,pigliapochi17,maxisch06,sgroi17,ceder11}
and, favoring integer occupations of atomic states,
stabilizes a charge-disproportionated ground states thus leading to a clear distinction between
2+ and 3+ Fe ions. In fact, as evident from the DFT+U$_{ave}$ results in Table \ref{fev}, 
consistently with previous work, half of the Fe ions in Li$_{0.5}$FePO$_4$ have an occupation
that closely resembles that of the (2+) Fe ions in LiFePO$_4$, 
the other half that of the (3+) Fe ions in FePO$_4$. 
When the Hubbard $U$ is computed self-consistently, within DFT+U, the roles of 2+ and 3+ are inverted: the 2+ Fe is not the one 
closest to the Li, as expected, but the one furthest apart.
However, as it will be discussed later, this is a spurious result, probably a consequence
of the system getting trapped in a local minimum of the energy.
Using the extended corrective functional DFT+U+V, 
the Fe ions closest to Li correctly recover their 2+ state. 

The results shown in Table \ref{fev} are based on a L\"owdin population analysis of Fe 3$d$ states, 
according to the definition given in Eq. \ref{occupij}.
That the numerical values are not in agreement with chemical practice (that would attribute 6 $d$ electrons
to Fe$^{2+}$ and 5 to Fe$^{3+}$) is related to both the mathematical definition of the
occupation matrices (not a univocal choice), to the physics of the system (e.g., some degree of hybridization
between Fe 3$d$ states and neighboring O 2$p$ states) and, in general, to other reasons discussed in
Ref.\cite{selloni11,selloni11_1}. However, it is important to remark that our Hubbard corrections
recover an electronic structure that is fully consistent with a charge-disproportionated 
state, with Fe$^{2+}$
and Fe$^{3+}$ ions described in full consistency with the end-members of the Li$_{x}$FePO$_4$ system.

Regarding the formation energy of the $x=0.5$ compound,
the GGA calculations 
return a negative value, that contradicts the lack of observation of stable 
phases at $x=0.5$. These results confirm the conclusions of previous work
\cite{zhou04} that already showed how this situation is due to the
over-stabilization of the incorrect metallic state obtained at intermediate Li content
(both the $x = 0$ and $x = 1$ materials are predicted to be insulators) which lowers the
total energy of the $x$ = 0.5 compound compared to the weighted average of the 
fully lithiated and delithiated end points. The incomplete localization of valence
electrons is also at the origin of the significant underestimation (-0.8 V) of the 
voltage: in fact, this quantity can be broadly related to the energy made
available to the system by each Fe ion when it is reduced from a 3+ to a 2+ oxidation state.
By promoting the localization of valence $d$ electrons DFT+U reverts this tendency and
produces a positive formation energy for $x=0.5$ and a voltage in much better
agreeement with the experimental value. However, the best results are obtained 
when the inter-site $V$ between Fe 3$d$ and O 2$p$ states is switched on, leading to
a predicted voltage which is less than 0.1 V off from the experimental value. 
In comparison, hybrid functionals, while improving electron localization and charge
disproportionation, generally result in average voltages of comparable quality \cite{urban16}, but
do not systematically improve errors on formation energies \cite{ceder11_1}.
\begin{figure}
\begin{center}
\subfigure[]{
\includegraphics[width=6.0cm,keepaspectratio,angle=-0]{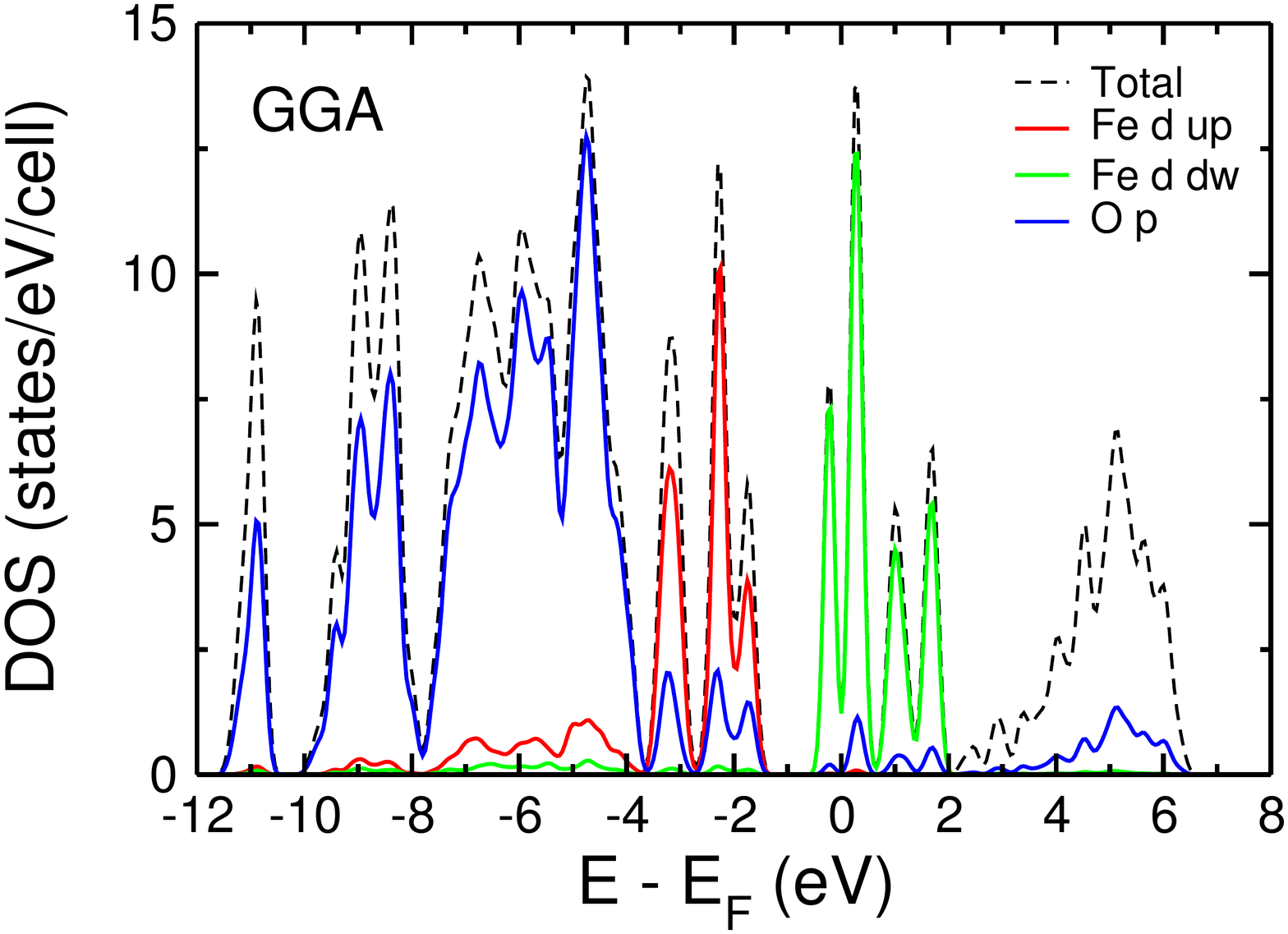}}
\subfigure[]{
\includegraphics[width=6.0cm,keepaspectratio,angle=-0]{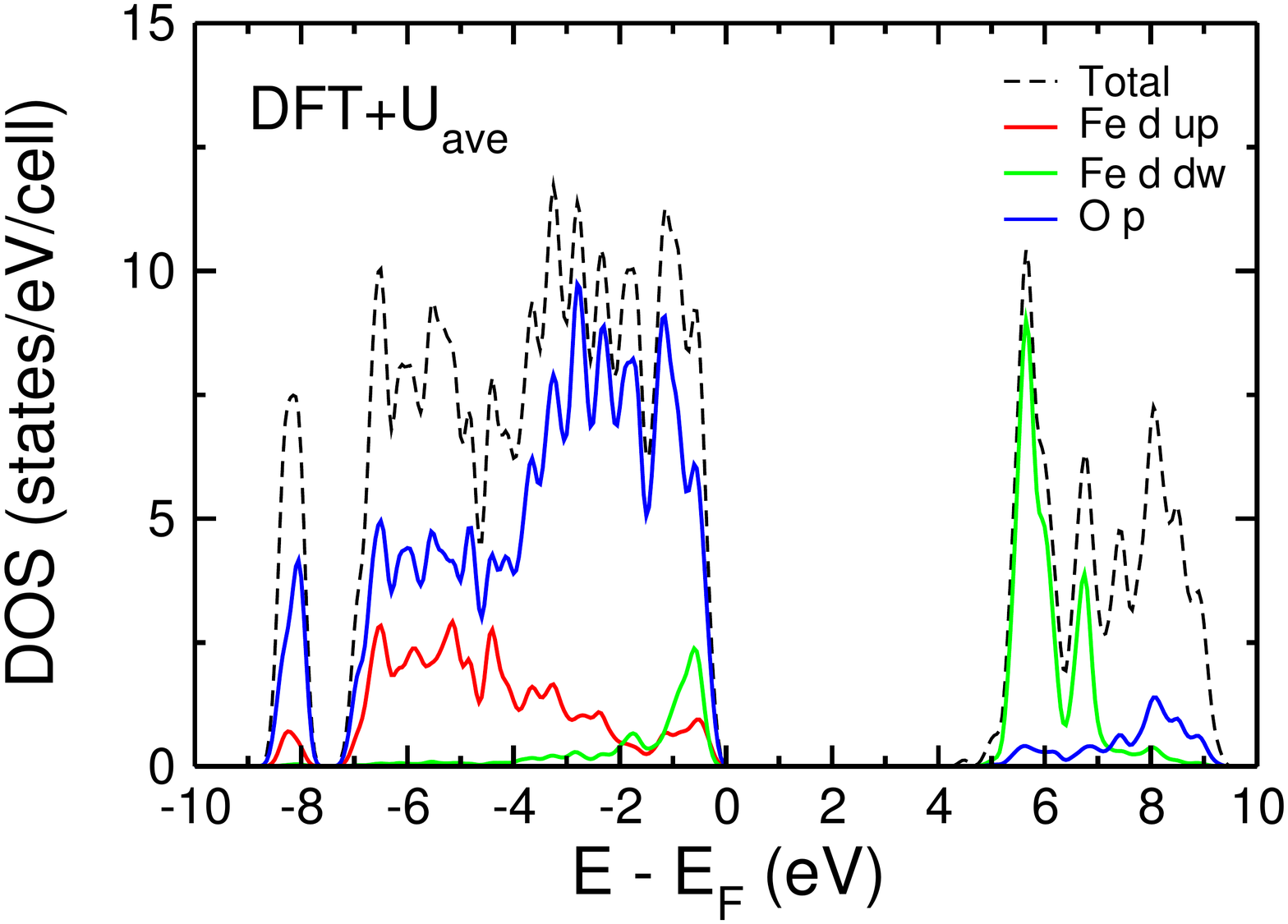}}
\subfigure[]{
\includegraphics[width=6.0cm,keepaspectratio,angle=-0]{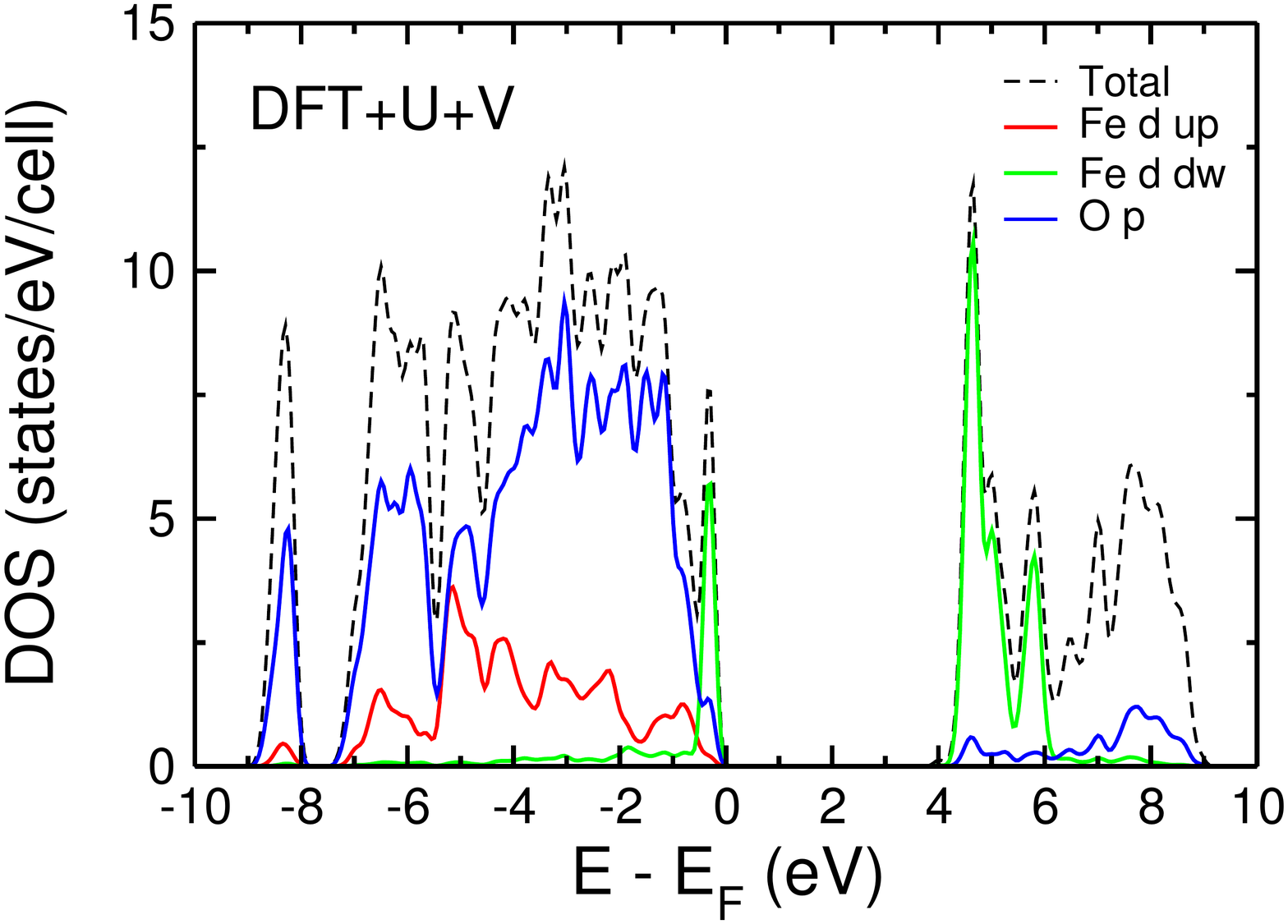}}
\caption{\label{lfpodos} (Color online) The density of states of LiFePO$_4$ obtained
with different approximations: (a) GGA (PBEsol); (b) DFT+U$_{ave}$; 
(c) DFT+U+V. 
In all the graphs the black dashed line represents the total density of state while
solid red, green and blue ones designate iron $d$ state spin up, iron $d$ state spin down and
oxygen $p$ states total contributions, respectively. All energies are referred to the Fermi level or
to the top of the valence band in the presence of a gap.}
\end{center}
\end{figure}
In order to better understand the effects on the ground-state electronic structure
of the various electronic interactions, we compare in Fig. \ref{lfpodos} the 
density of states (DOS) for the fully lithiated material, obtained with
the approximations discussed above. All graphs reported in the figure
show the total DOS (black dashed line) with the total contributions
from the transition-metal $d$ states (solid red lines for majority spin, green for minority)
and from (spin-unpolarized) oxygen $p$ states.
As evident from the figure, while the GGA (PBEsol) functional 
predicts a metallic behavior with the Fermi level of the system
crossing the minority-spin $d$ states, DFT+U is 
effective in opening a band gap (of about 4.5 eV if the average value of $U$
is used) in the Kohn-Sham spectrum of the material by separating a single 
occupied minority-spin $d$ state from the rest of its manifold
(Fe$^{2+}$ has nominally six $d$ electrons and in its highest spin configuration has 
five electrons in one spin state and the remaining one in the other).
DFT+U+V also predicts an insulating ground state; however the 
band gap is lower ($\approx$ 4 eV) than in DFT+U and closer to the experimental value
($\approx$ 3.7 eV). 
Another notable difference from the on-site only DFT+U is that the occupied
minority-spin $d$ state shows a lower contribution in the energy window dominated by oxygen
$p$ states and majority-spin $d$ states, while it dominates the energy window at the top of the valence 
manifold with a strong peak.

On a more technical level, it is fair to remark that the DFT+U$_{ave}$ and the DFT+U results presented
above can be expected to be slightly less accurate than the others, since the Hubbard
$U$ were obtained including the response of oxygen ions into the susceptibility matrices.
As mentioned at the end of section \ref{upvc}, this can lead to the under-screening
of the effective interaction parameter that results from the linear-response calculations.
In order to precisely assess the entity of this approximation DFT+U$_{ave}$ calculations
were repeated for Li$_{x}$FePO$_{4}$ with a Hubbard $U$ computed from a linear-response
procedure that only involved perturbing Fe ions. The value of the Hubbard
$U$ obtained this way (after averaging over the results of the three calculations
at $x$ = 0, 0.5 and 1), is 6.25 eV which, due to the additional screening from the 
oxygen $p$ states, is more than 0.5 eV lower than the one computed previously
(see Table \ref{uev}). In spite of this significant difference in the numerical value
of the effective parameter, however, the values of the formation energy of the $x$ = 0.5
compound and of the average voltage are 159 meV per formula unit and 3.96 V, respectively, 
which are very close with those reported in Table \ref{fev}. These quantities
thus seem to be robust with respect to the amount of screening included in the calculation of the effective
Hubbard parameters (although a more precise assessment should also take into account 
the optimization of the crystal structure). 
The same conclusions are corroborated by a further validation of the results presented so far that
was obtained by recomputing (self-consistently) the effective Hubbard parameters
from DFPT \cite{timrov18}. 
The details of these results are presented in the Supplementary Information \cite{suppinf19}.
An important outcome of these calculations is that the better control offered by DFPT on the convergence of the
Hubbard parameters and the lower numerical noise they are affected by in smaller cells
might help avoiding local minima (actually relatively common with Hubbard corrected functionals)
along the self-consistent calculation of interactions, electronic and crystal structures, as the one
found within DFT+U for Li$_{0.5}$FePO$_4$ and leading to the swapping of 2$^+$ and 3$^+$ Fe ions.

\subsubsection{Li-ion and magnetic configurations}

After a thorough discussion and a quantitative analysis of the results of various Hubbard
corrections a comparison is now performed between different possible configurations of the magnetic 
moments localized on the Fe ions and between two Li orderings already introduced at the beginning of
this section, alternating full and empty Li planes parallel, respectively, to the $yz$ and the $xz$
crystallographic planes. These comparative analyses will be aimed at establishing the most energetically
favorable configurations and at assessing, through the evaluation of energy differences, the possibility
for the system to visit higher energy (metastable) configurations, e.g. under the non equilibrium conditions
it experiences during the charge/discharge transients of the battery, especially in high currents 
regimes \cite{ceder11,ceder13,grey14}.
\begin{figure}
\begin{center}
\subfigure[]{
\includegraphics[width=7.0cm,keepaspectratio,angle=-0]{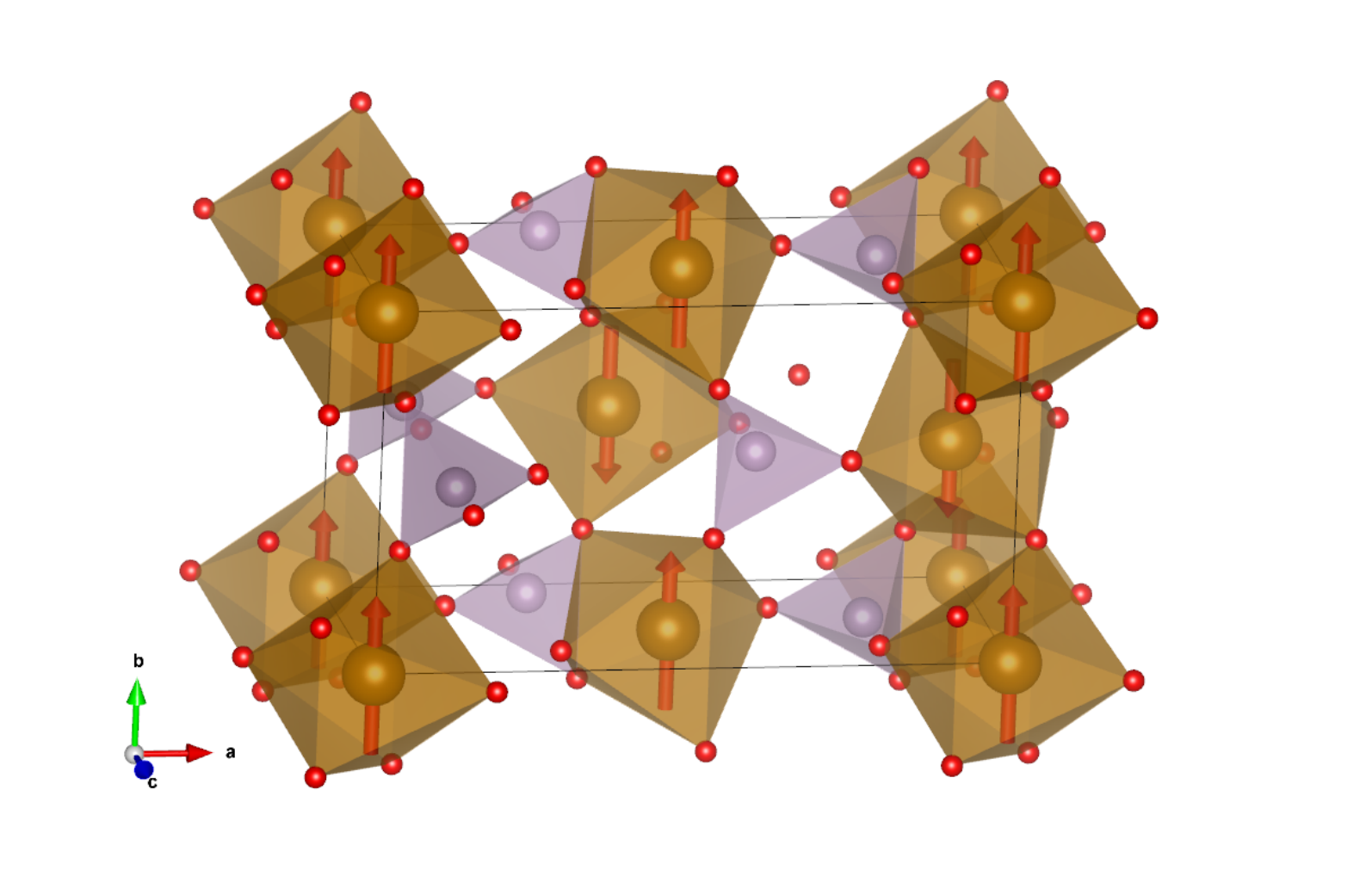}}
\subfigure[]{
\includegraphics[width=7.0cm,keepaspectratio,angle=-0]{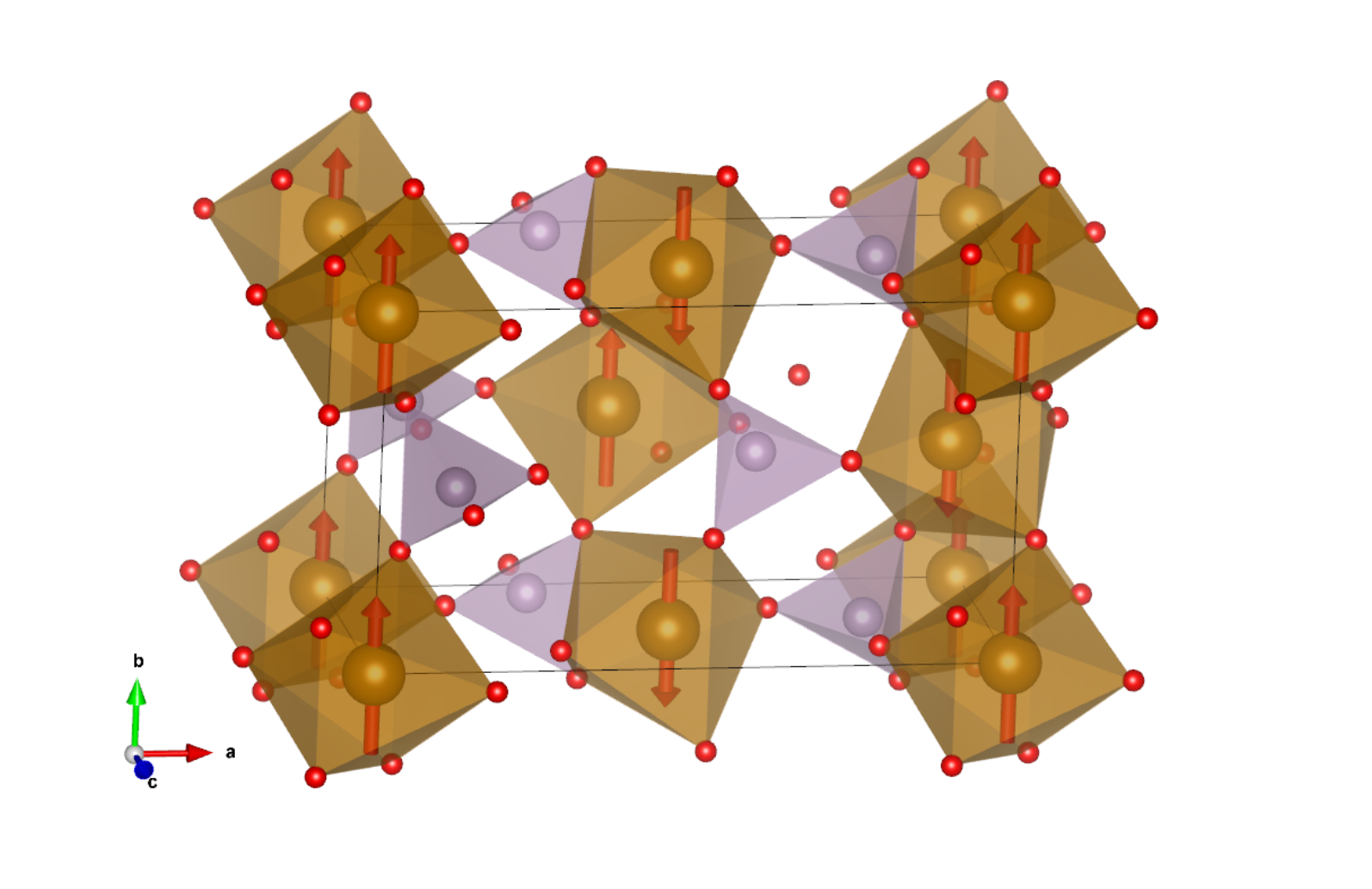}}
\subfigure[]{
\includegraphics[width=7.0cm,keepaspectratio,angle=-0]{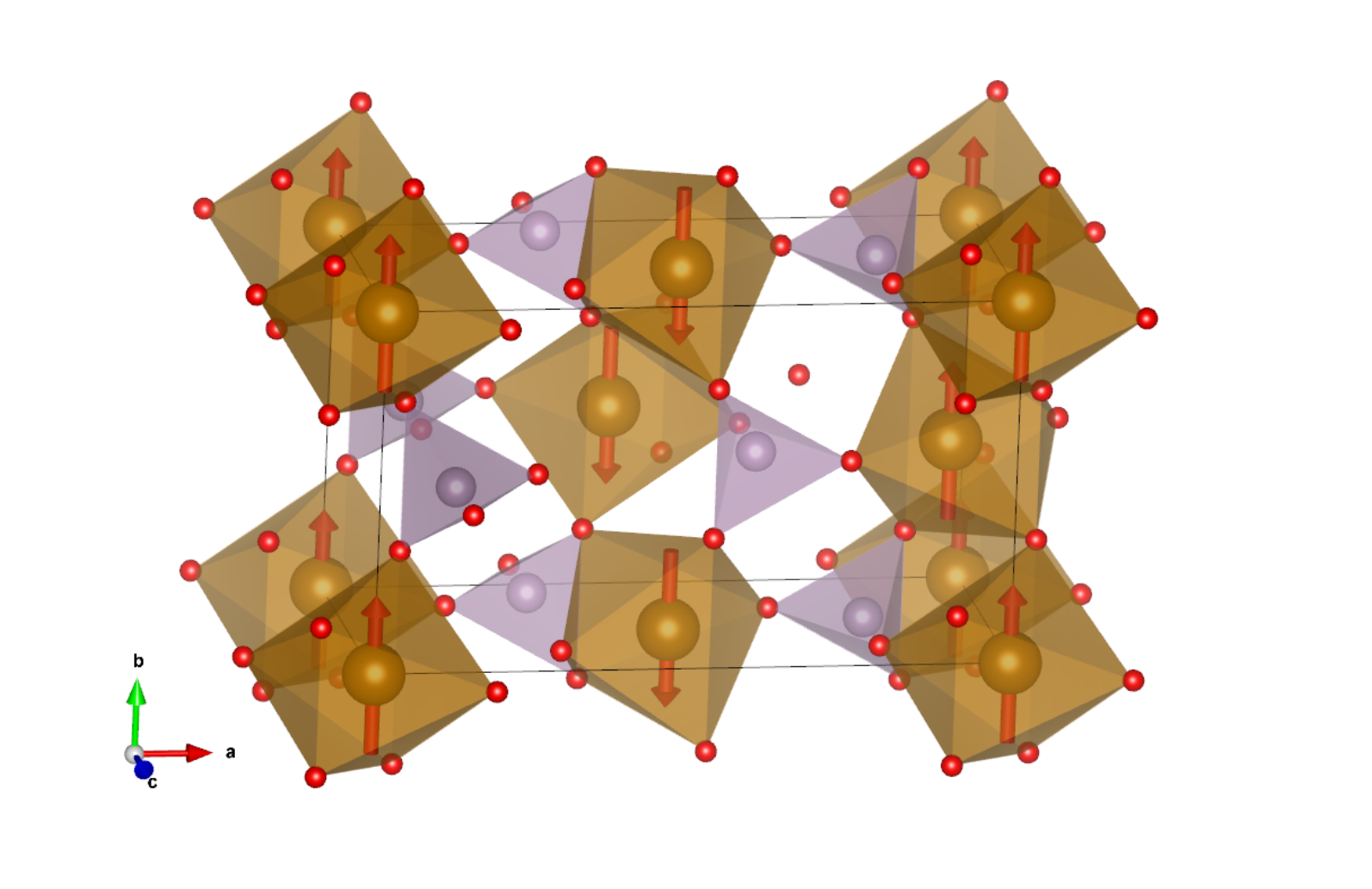}}
\subfigure[]{
\includegraphics[width=7.0cm,keepaspectratio,angle=-0]{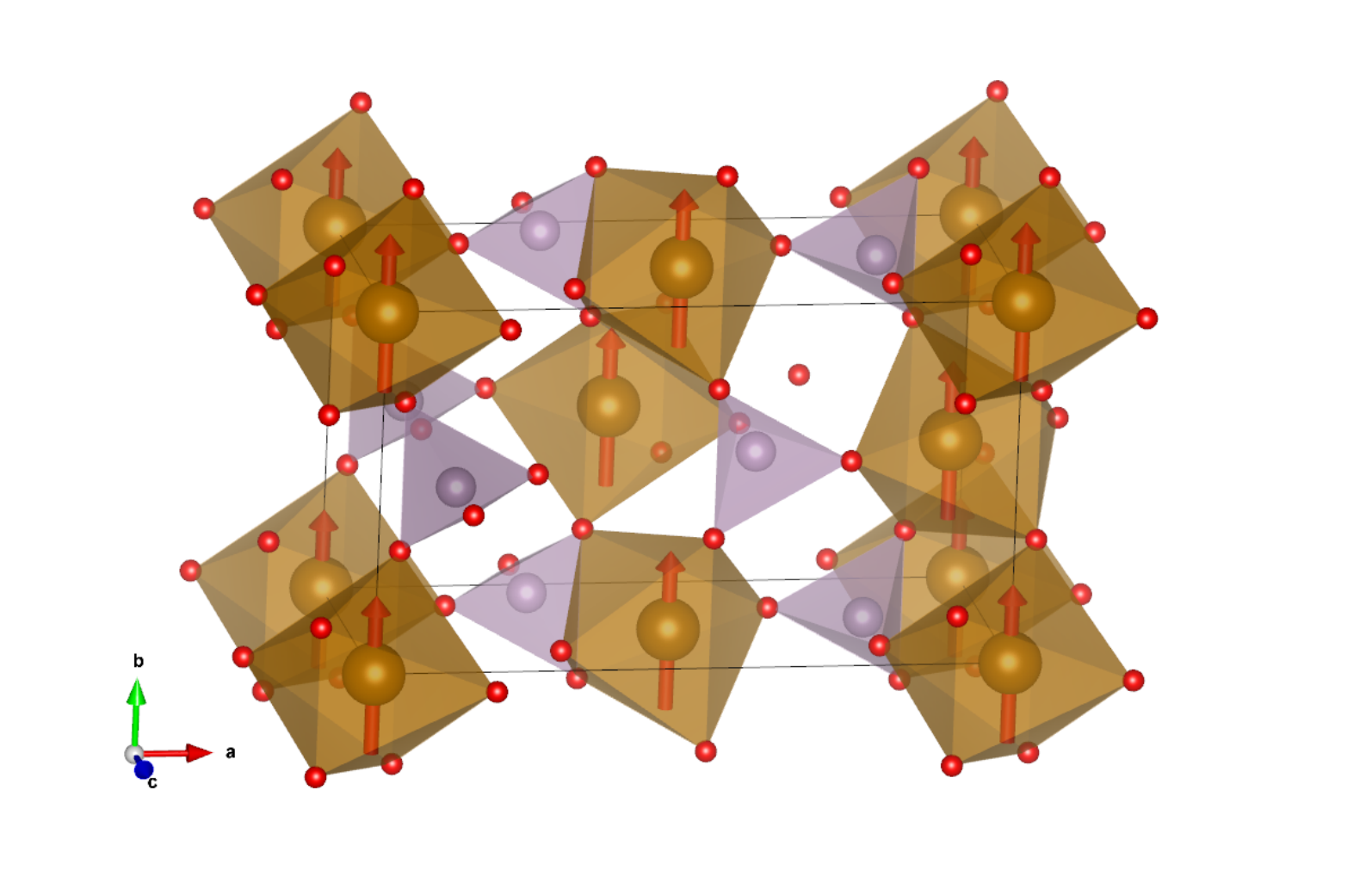}}
\caption{\label{afmconf} (Color online) The four magnetic configurations of
LiFePO$_4$ compared in the text: (a) AF$_1$, (b) AF$_2$, (c) AF$_3$, and (d) FM. Note that the 
Li atoms have been removed from the figures to improve clarity.} 
\end{center}
\end{figure}

\begin{center}
\begin{table}[h!]
\begin{tabular}{|c|c|c|c|c|c|c|c|}
\hline
\hline
 & $a$ & $b$ & $c$ & AF$_1$ & AF$_2$ & AF$_3$ & FM \\
\hline
Fe$_1$ & 0.00 & 0.00 & 0.00 & $\uparrow$ & $\uparrow$  &  $\uparrow$ &  $\uparrow$ \\
\hline
Fe$_2$ & 0.50 & 0.00 & 0.55 & $\uparrow$ & $\downarrow$ & $\downarrow$ & $\uparrow$\\
\hline
Fe$_3$ & 0.94 & 0.50 & 0.50 & $\downarrow$ & $\downarrow$ & $\uparrow$ & $\uparrow$ \\
\hline
Fe$_4$ & 0.44 & 0.50 & 0.05  & $\downarrow$ & $\uparrow$ & $\downarrow$ & $\uparrow$ \\
\hline
$\Delta E$ & \multicolumn{3}{c|}{} & 0.0 & 3.1 & 12.1 & 16.1 \\
\hline
\end{tabular}
\caption{The magnetic moments arrangements for the Fe ions in LiFePO$_4$ for all the four configurations
compared in this manuscript. The coordinates of the Fe ions are given in crystalline units and correspond
to the equilibrium structure obtained with DFT+U+V. 
$\Delta E$ represents the energy differences (in meV per formula unit) between them as obtained from the DFT+U+V
energy (see text for details), when PBEsol is used.}
\label{magconf}
\end{table}
\end{center}

The comparison between different magnetic configurations is performed only for the fully lithiated material
(LiFePO$_4$); however, similar conclusions are obtained also for FePO$_4$ and can presumably be extended to
all Li concentrations. 
Table \ref{magconf} provides a scheme illustrating the relative orientation of the magnetic moments on Fe ions 
(identified by their crystalline coordinates) in the three
antiferromagnetic and the ferromagnetic configurations, compatible with the primitive
(28-atoms) unit cell of the crystal. In its last row it also compares their energies as 
obtained from DFT+U+V calculations. 
The comparison is further eased by Fig. \ref{afmconf} which offers a visual representation of the unit cell
for the $x$ = 1 material (although Li ions were removed from the picture) in the four configurations considered. 
The arrows on the Fe ions indicate their magnetic moments.
Since all calculations were performed with a collinear spin functional
and with no spin-orbit coupling, the direction of the magnetic moments of the Fe ions is arbitrary and the arrows
in the figure point in a generic direction. 
In the first configuration (AF$_1$) the antiferromagnetic ground state is realized by ferromagnetic
(010) planes alternating with opposite magnetization. This spacial distribution of magnetic moments implies
that when Fe-O octahedra share one vertex oxygen their magnetization is antiparallel; conversely,
their moments are parallel if they are connected through a P-O tetrahedron.
In the second configuration (AF$_2$), instead, each Fe has a magnetic momentum antiparallel to that of all its first neighbors,
independently from how their octahedra are connected. Finally, the third antiferromagnetic configuration (AF$_3$)
consists of ferromagnetic (100) planes alternating with opposite magnetization.
As reported in Table \ref{magconf}, total energies of the various magnetic configurations are ordered 
as follows: E$_{AF1}$ $<$ E$_{AF2}$ $<$ E$_{AF3}$ $<$ E$_{FM}$. 
This ordering is consistent with previous work in literature \cite{rousse03,jena16} 
(at least for the ground and first-excited states) 
and is robust with respect the particular approximation being used, although
the energy differences between the various magnetic orders varies with the specific flavor of 
Hubbard correction in use. For example, for the PBEsol functional the energies of AF$_2$ and AF$_3$ are about 10.8, 19.3
meV/f.u. higher than that of AF$_1$, respectively.
The antiferromagnetic configuration that results as the ground state is the same adopted in the first part
of this section for all three Li concentrations considered. The comparison between various magnetic orders presented here
thus provides an a-posteriori justification for adopting it in the first place.
Relaxing the crystal structures turned out, in all cases, irrelevant for the ordering reported above
(the reported energies are the ones obtained from the optimized structures in all cases)
and, in many cases, changed very little the energy splittings between the various phases.
This result was obtained without adapting self-consistently the $U$ and $V$ interaction parameters and maintaining their
value equal to that of the groumd state AF$_1$. However, based on the relatively minor
changes in the crystal structure and on the weak dependence of the atomic charge on the 
relative spin arrangement, this approximation is expected to be quite good.
A further point in support of this procedure is the relative numerical irrelevance of the inter-site 
Fe-Fe terms of the response matrices (which are expected to be affected the most by a change in the relative
spin orientation) in determining the values of $U$'s and $V$'s.
In summary, independently from the functional adopted for the calculation and the particular flavor of
Hubbard correction, the energy difference between different magnetic configurations is very small
and largely irrelevant (in fact, one order of magnitude smaller) for the evaluation
of important quantities such as formation energies and average voltages.

\begin{figure}
\begin{center}
\subfigure[]{
\includegraphics[width=8.25cm,keepaspectratio,angle=-0]{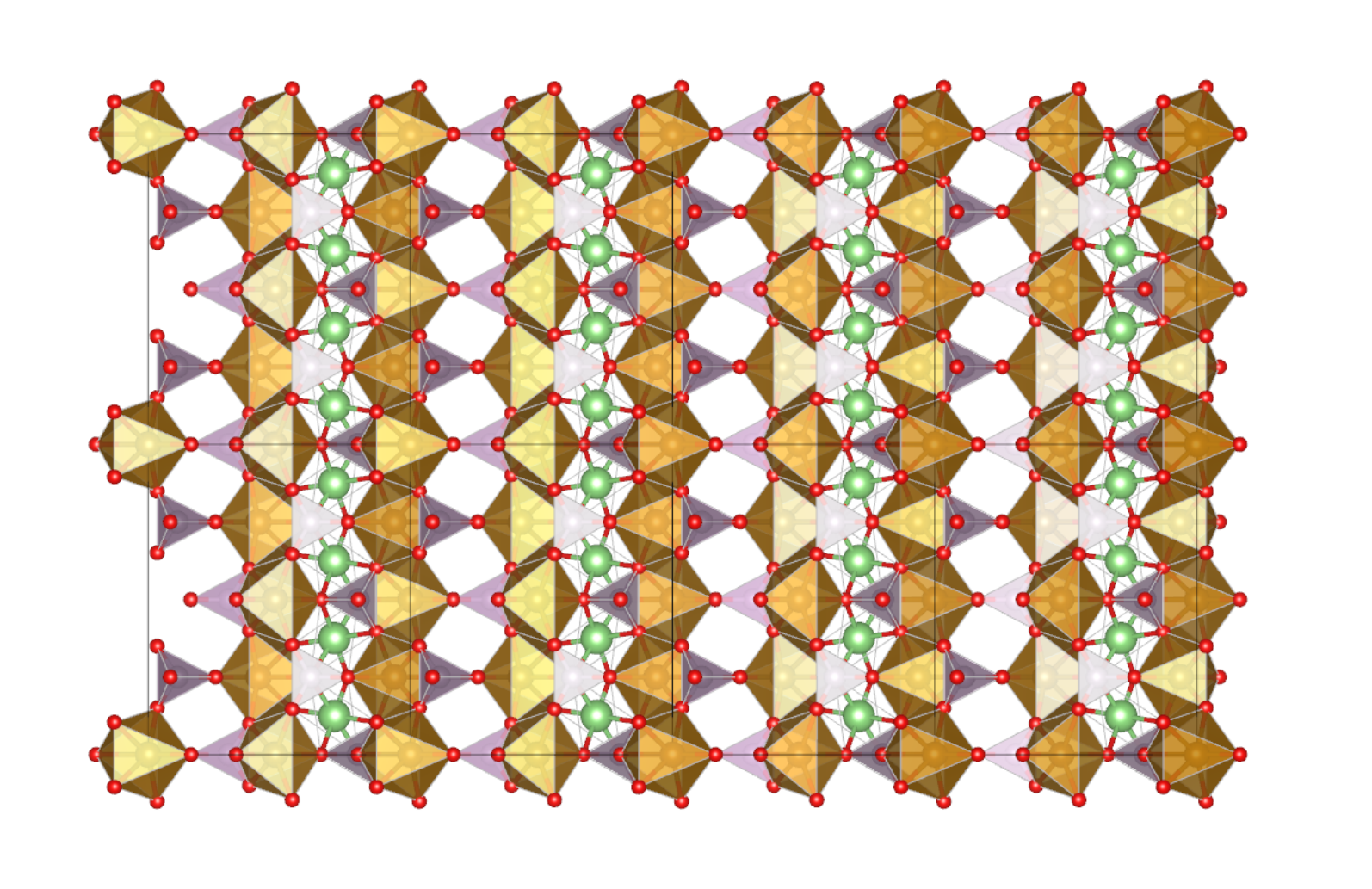}}
\subfigure[]{
\includegraphics[width=8.25cm,keepaspectratio,angle=-0]{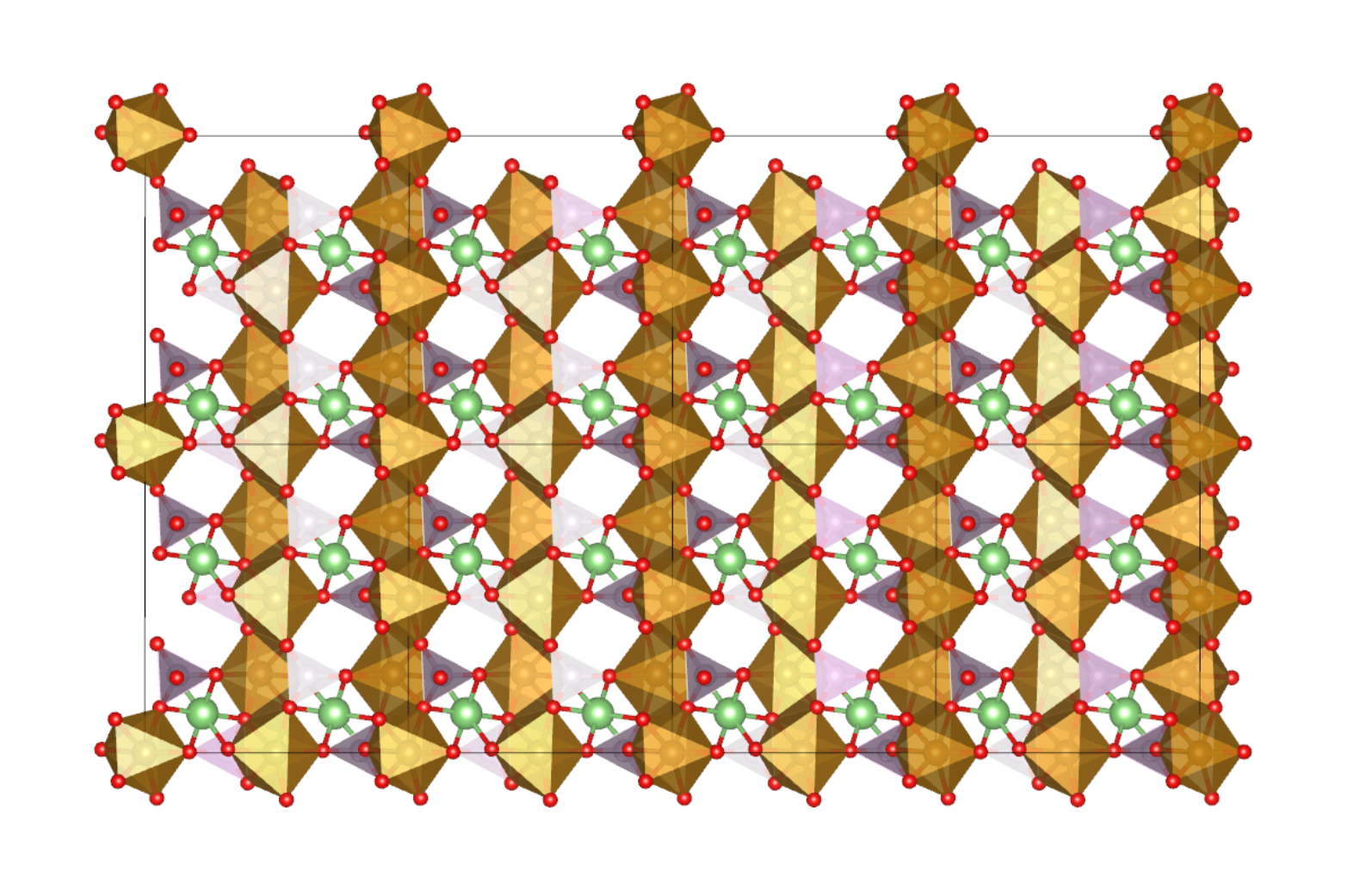}}
\caption{\label{l05fpo_12} The two Li configurations considered for Li$_{0.5}$FePO$_4$ with Li occupying 
alternate $yz$ (a) and $xz$ (b) planes. The two supercells present a view of the crystal along
the [001] direction.} 
\end{center}
\end{figure}
A second important comparative analysis is now developed between two Li configurations for
the half-lithiated Li$_{0.5}$FePO$_4$ material, shown in Fig. \ref{l05fpo_12}.
The first of them (Fig. \ref{l05fpo_12}(a)), with Li occupying alternate $yz$ 
planes (thus named ``$yz$" here) has been proposed to be forming in the surrounding of the interface between the $x=0$ and $x=1$
regions in small particles of the partially lithiated system,
during the charge/discharge transients (at least at low (de)lithiation rates) or  
after they have reached the equilibrium configuration 
(the observed samples were obtained by disassembling a half-charged battery, cycled at low rates) \cite{maier11,maier14_1}. 
According to these works this intermediate $x=0.5$ phase is stabilized
by the relaxation of the interfacial stress between lithiated and delithiated regions
(due to a milder lattice mismatch) until the energetic cost of its formation offsets the gain
in elastic energy.
The relaxation of interfacial elastic energy is also corroborated by the observation that
the thickness of this half-lithiated layer depends inversely on the size of the particle \cite{maier14_1}.
A previous computational study on this intermediate phase has highlighted instead its role in enhancing 
the motion of the-above mentioned interfaces during the charge and discharge of the battery through a 
significant reduction (compared to a sharp interface) of the kinetic barriers Li has to overcome during 
its diffusion throughout the material \cite{huang12}. 
The second of these half-lithiated phases, with Li occupying alternate $xz$ planes (and named ``$xz$"),
has been proposed to be one of the intermediate configurations the system visits when the (de)lithiation
process, at high charge/discharge rates (high overpotentials and currents), proceeds through 
non-equilibrium solid solutions of progressively lower/higher Li content \cite{ceder11,ceder13}.
This mechanism is expected to be viable even at relatively low temperatures due to the low formation energy
of a plethora of partially lithiated phases with respect to the end members.
Based on the results presented previously,
our comparative analysis on these crystals is based on DFT+U+V calculations only, with $U$ and $V$ 
evaluated self-consistently with both the electronic and crystal structures as explained
in section \ref{tecdet}, for both configurations. The main result of our calculations 
concerns the total energy of the two $x=0.5$ phases. 
We obtain the second phase to be lower in energy than the first by about 65 meV per formula unit and, 
with a formation energy of 42.5 meV per formula unit, it results in being relatively more accessible for a 
system at finite temperatures. 
While significantly lower than that of the $yz$ configuration, this formation energy is however
somewhat higher than 
that reported in Refs. \cite{ceder11,ceder13}, obtained from Monte Carlo simulations. 
It is interesting to note that the different position of Li ions also induces a redistribution
of valence electrons on the transition-metal ions; in particular, between the two configurations, 
the Fe ions swap their oxidation state (the ones that are in a 2+ state in the first become 3+ 
in the second, and viceversa). Consistently with chemical intuition and with the localized
nature of $d$ electrons, the Fe ions that are closer to the Li
result in both cases in the 2+ oxidation state, thus confirming that Li motion correlates
with the diffusion of localized electronic charges in their neighborood. 
As highlighted previously, this correlation between the position of Li ions and electrons 
is quite hard to capture with approximate exchange-correlation functionals, especially for 
the second configuration in which the difference between the Fe-Li distances for the 2+ and 3+ ions is significantly smaller.  
The use of the extended Hubbard correction, with a consistent calculation of the
interaction parameters, produces instead a consistent description of the two valence states of Fe ions
(both categories of Fe ions having practically the same atomic occupation in the two configurations)
while polarizing the crystal structure around Fe ions consistently with their valence. 
It is interesting to note that also the Hubbard $U$'s assume values that are consistent between
the two Li configurations for ions in the same oxidation state, with those for the second
structure resulting in values of $U$ of 4.96 and 4.86 eV for the 3+ and 2+ Fe ions, respectively, in excellent
agreement with those shown in Table \ref{uev}. The values of the inter-site interaction
parameters, instead, do not show the same agreement between the two configurations (those
for the second structure result in the 0.19 - 0.54 eV and 0.15 - 0.59 eV ranges for the
3+ and 2+ ions, respectively), probably reflecting a different distortion of the oxygen
octahedra coordinating the Fe centers.

\subsection{Li$_x$MnPO$_4$}
\subsubsection{Electronic structure and energetics}
The potential of Li$_x$MnPO$_4$ (LMPO) as a possible cathode material
for Li-ion batteries was already recognized in the seminal work by Padhi et al. \cite{padhi97} 
that has first promoted phospho-olivines as positive electrode materials. 
Compared to the isostructural LFPO, LMPO presents 
a higher voltage (4.1 V vs Li/Li$^+$) which is quite an appealing feature
in perspective of its use in higher energy density cathodes, 
still within the breakdown potential of most liquid electrolytes currently available.
Unfortunately, LMPO is also plagued by several problems that have so far 
prevented its deployment in actual devices. Among these, particularly harmful are
a lower chemical stability (especially of the delithiated phase)
that makes the material prone to parasitic reactions with consequent rapid loss
of capacity; possibly less safe and more difficult to synthesize using the same techniques
successful for LFPO\cite{chen10,madhavi13,herrera15}; a larger volume/lattice mismatch
between lithiated and de-lithiated phases that increases the chances of permanent structural 
defects at interfaces and makes the material loose its cyclability more rapidly \cite{lei10,liu10};
a difficult deposition of C on the surface of nanostructured LMPO samples, needed to improve their 
electronic conductivity \cite{madhavi13,graetz11,julien12};
the Jahn-Teller (JT) activity of Mn$^{3+}$ ions (in the delithiated
phase), promoting quite significant structural distortions
in their local environment (in particular of the oxygen cage around them) and abrupt changes also
in the electronic structure that may impact negatively the performance of the material.
The most serious problem for using LMPO in actual devices is represented, however, by
its low ionic and electronic conductivities that, while further compromised by 
passivation events during the charge and discharge of the battery or by lattice disorder
(with Mn$^{2+}$ ions possibly obstructing the Li diffusion channels) \cite{liu10}, are generally
attributed to the significant lattice polarization around injected charges (leading to the
formation of polarons)
that are caused or enhanced by the JT activity of Mn$^{3+}$ ions and result in sluggish motion of the
charge carriers (i.e., high effective masses or higher kinetic barriers for the hopping to neighbor sites)
\cite{yamada06,ceder11_1}. 
The necessity to find a way around these problems and to define viable strategies to employ LMPO
in actual cathodes of Li batteries has stimulated a quite lively research activity on this 
material (summarized, e.g., in Ref. \cite{herrera15})
that has encompassed the precise characterization of its structural, magnetic and
electrochemical properties \cite{yamada06,martha09}, but also the development of new fabrication
techniques that could improve its performance as cathode 
material \cite{liu10,pivko12,taniguchi11,tohda02,graetz11,julien12,xiao10}.
Several computational studies have also been performed on this system, generally
focusing on the electronic, magnetic, structural and vibrational properties of LMPO \cite{goel14}. 
A particular attention in this context has been devoted to the study of
the local distortion around the (possibly JT-active) 3+ Mn ions and to its impact on 
the ionic and electronic conductivities \cite{lei10,whittingham13}, occasionally addressed
quantitatively through the evaluation of relevant kinetic barriers \cite{ceder11_1}.

In this work a computational study of this material is presented that focuses on
ground state electronic, magnetic and structural properties. The same Li concentrations
examined for LFPO are also considered for LFMO and for each of them a ground
state consistent with the choice of the Hubbard flavor and the value of the
interaction parameters is computed. A similar comparative analysis between
the three materials is also performed discussing the crystal structures (with
a particular focus on the local environment of Mn ions), the number of electrons
on their $d$ orbitals (directly related to their oxidation state) and the total energies.
The latter will be employed for the calculation of the formation energy of the $x = 0.5$ system
and of the average voltage vs Li/Li$^+$.
Because of the role played by the hybridization between Mn and O in determining 
the local distortion of the crystal and its transport properties,
LMPO will represent a particularly
significant test case for the extended Hubbard functional
discussed in this work and will highlight the importance of the 
inter-site interactions to capture the properties of systems where valence
electrons do not completely localize on atomic states.
As in the case of LFPO, also for this system a comparative analysis is performed
between various antiferromagnetic configurations of the $x=1$ compound and between
different possible ordered phases of the half-lithiated material. 
Given the similarity between the two systems, the same Li and magnetic configurations of the Fe-based
material are considered. 
The results are discussed in the last part of this section, while its first part
concerns the AF1 configuration and the $x=0.5$ structure with 
Li filling alternating $yz$ planes. 

As mentioned in section \ref{tecdet}, the computational study on this system was performed using a PAW pseudopotential
\cite{blochl94} which also included Mn 3$s$ and 3$p$ states into the valence manifold. 
These calculations required kinetic energy cutoffs of 100 and 400 Ry for the 
plane-wave expansion of the electronic wave functions and charge density,
respectively. Brillouin zone integrations were performed using the same
2$\times$4$\times$4 special k-point grid also adopted for LFPO.
\begin{widetext}
\begin{center}
\begin{table}[h!]
\begin{tabular}{|c|c|c|c|c|}
\hline
\hline
 & & LiMnPO$_4$ & \multicolumn{1}{c|}{Li$_{0.5}$MnPO$_4$} & \multicolumn{1}{c|}{MnPO$_4$}  \\
\hline
\multirow{1}{*}{DFT+U$_{ave}$} & U$_{Mn}$ & \multicolumn{3}{c|}{5.66}  \\
\hline
\multirow{1}{*}{DFT+U} & U$_{Mn}$ & 5.28 & 7.48/8.28  & 8.20 \\
\hline
\multirow{2}{*}{DFT+U+V} & U$_{Mn}$ & 4.44 & 4.96 / 6.27 & 6.28\\
& V$_{Mn-O}$ & 0.55 - 1.20 & 0.24 - 1.62 / 0.52 - 1.26 & 0.56 - 1.39\\
\hline
\end{tabular}
\caption{The values of $U$s and $V$s (in eV) obtained for Li$_{x}$MnPO$_4$, $x = 0$, 0.5, 1, 
computed within various flavors of the Hubbard correction (see text).
The ranges of values reported for the $V$ parameters refer to the first
coordination shell (their values vary with the M-O distance).
For Li$_{0.5}$MnPO$_4$ the two sets of values refer to the 2+ and 3+ Mn ions
respectively.}
\label{uevmn}
\end{table}
\end{center}
\end{widetext}
Table \ref{uevmn} reports the values of the effective $U$'s and $V$'s obtained for 
the Li$_x$MnPO$_4$ system ($x = 0$, 0.5, 1) using the self-consistent
linear-response approach discussed in section \ref{upvc}. 
Similarly to the case of LFPO even in this case the values
of the Hubbard parameters referring to Mn ions in the same oxidation state
can change quite significantly with Li content, 
especially for 2+ Mn ions. 
At the same time, the effective interactions change their values also
in dependence of the specific Hubbard functional adopted in their self-consistent evaluation.
For example, the value of $U$ varies quite significantly 
depending on whether the inter-site $V$ is used in the total energy calculations or not, and 
on whether structural relaxations are involved in the procedure.
As in the case of LFPO, the value of Hubbard parameters vary quite significantly with the oxidation
state. It is important to remark that the different values of $U$ and $V$ for the 2+ and 3+ ions in the 
half-lithiated material were obtained as a result of the linear-response approach discussed in section \ref{upvc}.
Their differenciation was not enforced in any way, although it is certainly the effect of a different 
crystal environment around 2+ and 3+ sites.
All these observations confirm the scarce portability of the Hubbard parameters and 
the necessity to compute them consistently for the system of interest and with the same approximate
functional (exchange-correlation and Hubbard correction) that is used in the calculations.
\begin{widetext}
\begin{center}
\begin{table}[h!]
\begin{tabular}{|c|c|c|c|c|c|}\hline
\hline
 & & \multicolumn{1}{|c|}{GGA} & DFT+U & \multicolumn{1}{c|
}{DFT+U+V} & \multicolumn{1}{c|}{Exp} \\
\hline
\multirow{3}{*}{LiMnPO$_4$} &
 a & 19.62 & 19.94 & 19.79 & 19.76$^a$, 19.71$^b$ \\
& b/a & 0.58 & 0.58 & 0.58 & 0.58$^{a,b}$\\
& c/a & 0.45 & 0.45 & 0.45 & 0.45$^{a,b}$\\
\hline
\multirow{6}{*}{Li$_{0.5}$MnPO$_4$} &
a & 18.74 & 19.39 & 19.19 & \multirow{6}{*}{n/a} \\
& b/a & 0.61 & 0.61  & 0.60 & \\
& c/a & 0.48 & 0.47 & 0.47 & \\
& $\alpha$ & 90.02 & 90.0 & 89.78 & \\
& $\beta$ & 88.44 & 86.93 & 86.03 & \\
& $\gamma$ & 89.95 & 90.0 & 93.35 & \\
\hline
\multirow{3}{*}{MnPO$_4$} &
a & 18.38 & 18.94 & 18.75 & 18.31$^b$ \\
& b/a & 0.61 & 0.62 & 0.61 & 0.61$^b$ \\
& c/a & 0.50 & 0.50 & 0.51 &  0.49$^b$\\
\hline
\end{tabular}
\caption{The equilibrium lattice parameters (in bohr) and the angles between the primitive cell 
vectors (degrees) of Li$_x$MnPO$_4$ ($x = 0$, 0.5, 1) computed with DFT (GGA at the PBEsol level) and with
the Hubbard +U and +U+V corrections,
and compared with available experimental values (the superscripts $a$ and $b$ indicate 
Ref. \cite{xiao10}, and Ref. \cite{lei10}, respectively).}
\label{lpmn}
\end{table}
\end{center}
\end{widetext}
Table \ref{lpmn} allows to compare the equilibrium crystal structure obtained
for this system within various approximations. 
In general, given the overall agreement on $b/a$ and $c/a$, it is possible to ascribe
the mismatch with the experimental results to an almost uniform linear scale factor.
For the fully lithiated compound GGA slightly underestimates (up to about 2\%) the lattice 
parameters with respect to their experimental value. The self-consistent DFT+U over-corrects 
this tendency resulting in equilibrium 
cell axes that are longer than predicted by experiments. 
When the inter-site interaction is turned on the general agreement between the 
computed and the experimental crystal structure is significantly improved with
differences that are within a fraction percent for all the structural parameters.
The same trends just discussed for $x = 1$ can be observed for the half-lithiated
and delithiated compounds, with DFT+U+V correcting the expansion of the unit
cell stabilized when only the on-site $U$ is used and predicting values for the 
lattice parameters that are intermediate between GGA and DFT+U.
However, for $x=0$ the best agreement with experimental results is obtained from GGA
(slightly overestimating lattice parameters), 
while DFT+U+V shows a mismatch with the experimental cell parameters in the 2 - 3\% range.
This result, at variance with what was obtained for FPO (see Table \ref{lattice_lfpo}),
is somewhat surprising in light of the overall improvement obtained with DFT+U+V on the
electronic structure and energetics (average voltage) of the material (see below).
The half-lithiated compound, modeled by the same ``staged" arrangement of Li atoms
also used for LFPO, deserves special attention. In fact, 
as evident from Table \ref{lpmn}, the lower symmetry of the crystal structure
can lead to significant distortions of the unit cell. While GGA and DFT+U stabilize an
effectively monoclinic cell (with only the $\beta$ angle deviating significantly from
90$^{\circ}$), the use of inter-site interactions $V$ in the corrective 
functionals leads to an equilibrium triclinic unit cell with all the
angles between cell vectors deviating from 90$^{\circ}$. 
\begin{widetext}
\begin{center}
\begin{table}[h!]
\begin{tabular}{|c|c|c|c|c|c|}
\hline
\hline
  & Mn$^{2+}$ ($x = 1$) & Mn$^{2+}$/Mn$^{3+}$ ($x=0.5$) & Mn$^{3+}$ ($x = 0$) & F. E. (meV/f.u.) & Voltage (V) \\
\hline
GGA & 5.30 & 5.19/5.17 & 5.11 & 63 & 2.82 \\
\hline
DFT+U$_{ave}$ & 5.19 & 5.11/5.05 & 4.96 & 212 & 4.31 \\
\hline
DFT+U & 5.18 & 5.11/5.08 & 4.98 & 161 & 5.14 \\
\hline
DFT+U+V & 5.23 & 5.22/4.99 & 4.99 & 206 & 4.15 \\
\hline
Exp & & & & \textgreater~0 & $\sim 4.1$ \\
\hline
\end{tabular}
\caption{Atomic occupations of $d$ states, formation energy, and average voltages
for Li$_x$MnPO$_4$, computed with different methods and compared with available experimental data. As also
noted in Table \ref{fev} for LFPO, DFT+U+V improves the agreement 
of the computed voltage with the experimental data and is the only approach to predict charge disproportionation
with an accurate description of atomic occupations in the mixed-valence ground state for $x = 0.5$.}
\label{fevmn}
\end{table}
\end{center}
\end{widetext}
Table \ref{fevmn} reports the total occupation of Mn atoms in the three Mn compounds
considered and compares the average voltages and the formation energies
for the half-lithiated crystal obtained from different Hubbard corrections.
At variance with LFPO, the DFT+U with on-site only corrections is
not effective in capturing the disproportionation of the L\"owdin charges in the half-lithiated
compound and the difference between the occupation of 2+ and 3+ Mn ions
in this case is only marginally larger than that predicted by GGA. When
using the inter-site $V$, instead, the occupations of 2+ and 3+ ions are obtained
in precise consistency with the fully lithiated (2+) and delithiated (3+) 
cases, showing the effectiveness of this corrective functional in capturing 
electronic localization in presence of significant hybridization. 
Regarding the formation energies, while they result positive in all the approximations considered,
their values increase when a Hubbard correction (of any type) is used, probably due to 
the destabilization of the metallic ground state that is, instead, obtained from GGA calculations
for the half-lithiated material. 
For the average voltage, however, while GGA significantly underestimates
experiments, DFT+U overestimates them. 
The best agreement with experimental measurements of this quantity
is obtained again within DFT+U+V.
\begin{figure}
\begin{center}
\vspace{-0.in}
\subfigure[]{
\includegraphics[width=6.0cm,keepaspectratio,angle=-0]{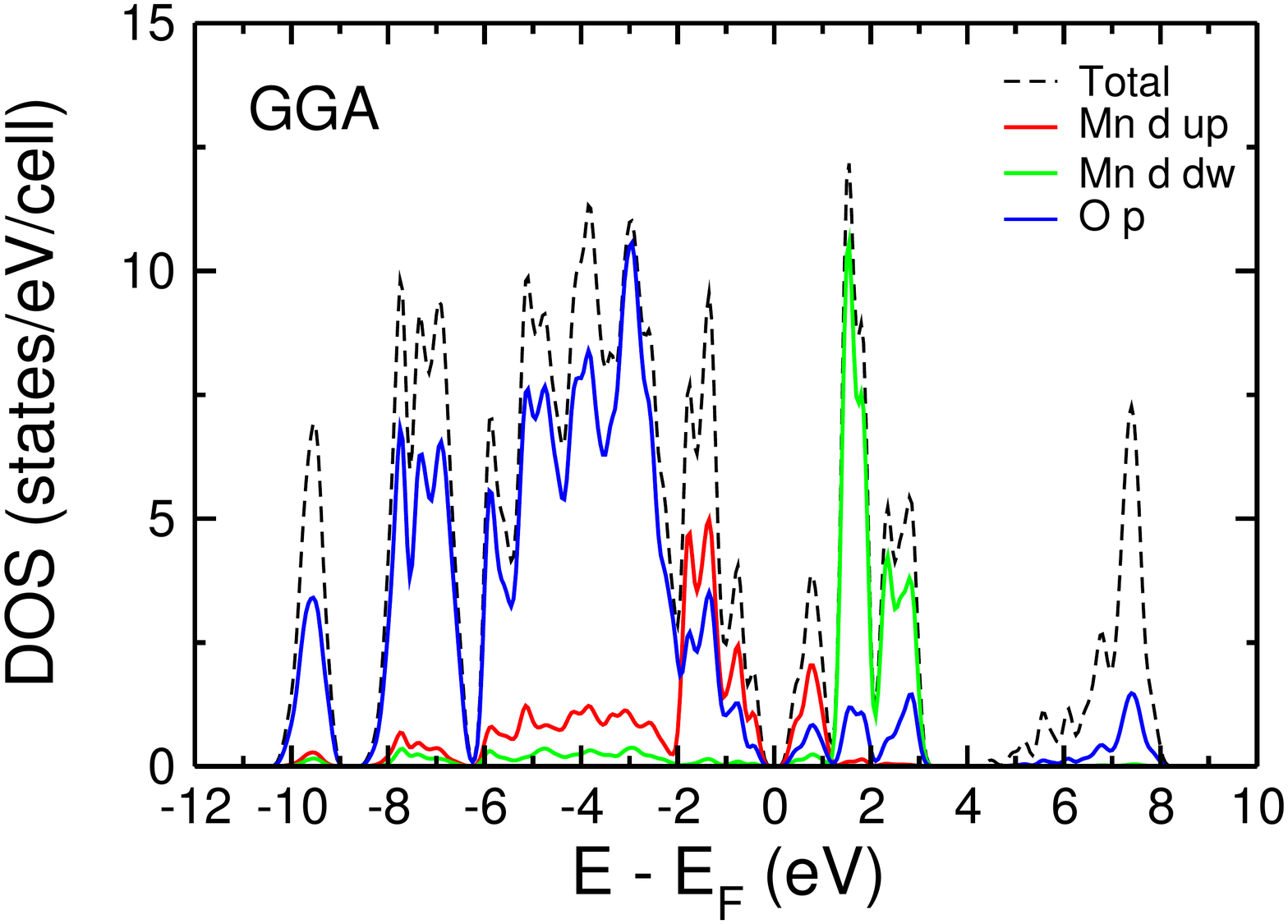}}
\vspace{-0.in}
\subfigure[]{
\includegraphics[width=6.0cm,keepaspectratio,angle=-0]{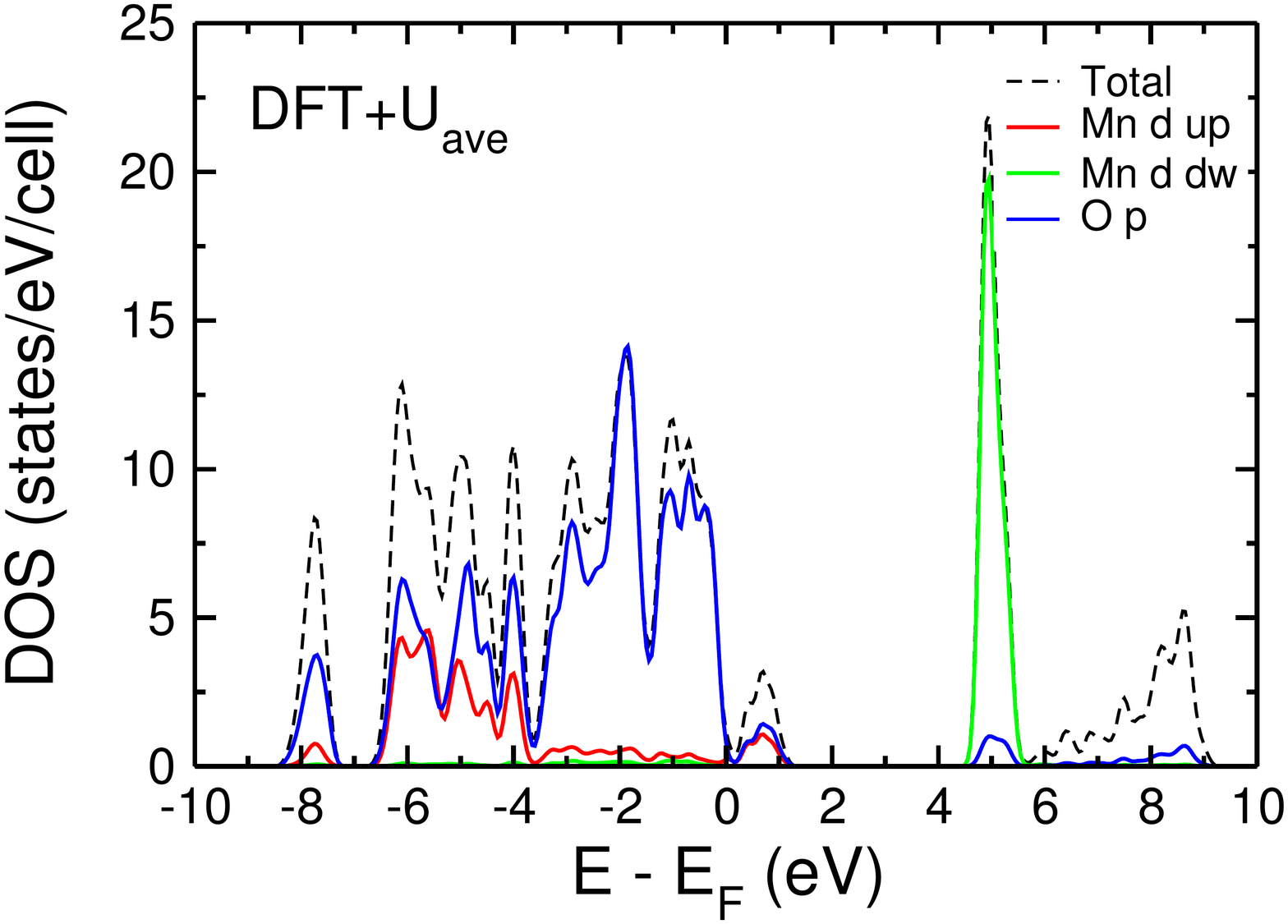}}
\vspace{-0.in}
\subfigure[]{
\includegraphics[width=6.0cm,keepaspectratio,angle=-0]{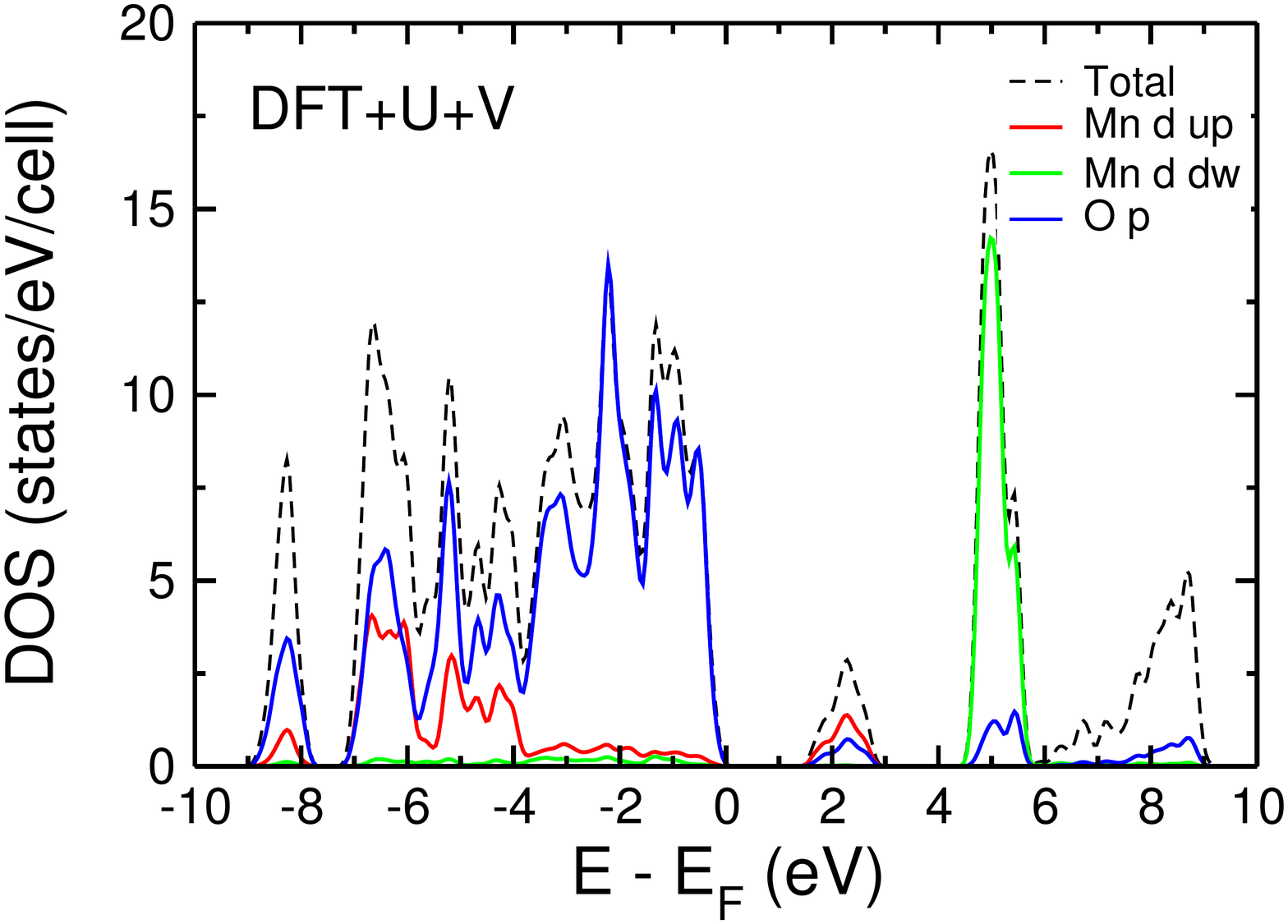}}
\vspace{-0.in}
\caption{\label{lmpodos} (Color online) The density of states of MnPO$_4$ obtained
with different approximations: GGA (PBEsol) (a); DFT+U$_{ave}$ (b);
DFT+U+V (c).
In all the graphs the black dashed line represents the total density of state while
solid red, green and blue ones designate manganese $d$ state spin up, manganese $d$ state 
spin down and oxygen $p$ states total contributions. All energies are referred to the Fermi 
level or to the top of the valence band in presence of a gap.}
\end{center}
\end{figure}
As in the case of LiFePO$_4$, in order to fully appreciate the effectiveness of the various
Hubbard corrections discussed above, a comparative study between the density of states obtained from
these approximations is proposed for one of the materials in the LMPO family, namely MnPO$_4$.
The de-lithiated member of the Mn olivine system has not been very much studied, probably
due to its poor thermal stability (its tendency to decompose and to release oxygen), actually still 
under investigation (see, e.g., Ref. \cite{huang14} and references therein).
Its computational characterization has been also quite sporadic with 
studies predicting either half-metallic \cite{dai14} or semiconducting \cite{zhou04_3} behavior.
The results of this work are shown in Fig. \ref{lmpodos} with the same color convention already used
in Fig. \ref {lfpodos} for LFPO.
For the correct interpretation of these results it is useful to keep in mind
that Mn ions are nominally in a 3+ oxidation state in this compound with four electrons in their
$d$ shells. Because of their high-spin configuration, the presence of a hole in 
their $d$ manifold should result in a majority-spin $d$ state moved above (and separated
by an energy gap from) the top of the valence band.
As clearly shown in Fig. \ref{lmpodos} this is only obtained when a finite inter-site
interaction $V$ is used in the (extended) Hubbard functional [Fig. \ref{lmpodos}(c)].
The GGA (PBEsol) functional does in fact yield a non-metallic ground state (the DOS of the system being 0
at the Fermi level). However the band gap is
quite small, probably due to the well known tendency of this functional to over-stabilize partially 
filled bands and delocalized distributions of electrons, and to  
its inability to completely emptying the highest energy majority-spin and the
minority-spin $d$ states. This result is in agreement with Ref. \cite{zhou04_3}.
In contrast with that work, instead, our DFT+U calculations
do not predict a semiconducting ground state and only result in an
upward shift of the energy of minority-spin $d$ states that are
minimally overlapping with oxygen $p$ states. The highest energy majority-spin $d$ state of Mn, instead,
probably because of its mixed $d$-$p$ character (as suggested by the presence of both contributions
in the peak right above the Fermi level), is predicted to be contiguous with the top of the valence band.
The use of the inter-site interaction $V$ qualitatively changes these results and, thanks to the more flexible
expression of the corrective functional, successfully localizes the hole on the
antibonding state formed by a majority-spin Mn $d$ and an O $p$ state, 
pushing the energy of the corresponding peak of the DOS
to higher energy and opening a finite energy gap (of about 1.1 - 1.3 eV)
from the top of the valence band.
From the comparison of DFT+U and DFT+U+V results it is important to note that 
the inter-site interaction mostly affects the top of the 
valence band, where the gap opens. In other regions of the energy spectrum its effect seem
to be minor, even if the overlap between Mn $d$ and O $p$ states is significant. 
This insensitivity is probably the result of two opposite factors that compensate each other: 
a smaller on-site Hubbard $U$ and a finite inter-site $V$ (see table \ref{uevmn}).

The hybridization-driven band-gap opening and hole localization described above 
for MPO is analogous to that discussed in Ref. \cite{ceder11_1} focusing on
the localization of an electron/hole polaron in LMPO. In that work DFT+U was found unable to 
achieve the full localization of the defect charge and is reported to predict a metallic 
behavior under all conditions. In order to predict a localized extra charge in the
material a hybrid functional was used, whose main effect
is to push the energy of the unoccupied majority-spin $d$ state of one Mn ion into the
band gap of the material. The results described above, although obtained from 
charge-neutral calculations, suggest that DFT+U+V could 
achieve a similar result. 

As in the case of LFPO, the results presented above are validated by a series of self-consistent calculations
including structural optimizations and Hubbard parameters evaluations from the DFPT\cite{timrov18}.
The results of these calculations are detailed in the Supplementary Information \cite{suppinf19} 
that develops a comparative analysis.
Overall, DFT+U+V calculations employing DFPT to compute the Hubbard parameters confirm the results discussed above
both quantitatively (the average voltage is 4.21 V) and qualitatively, highlighting their convergence and robustness.
A somewhat wider variation is instead noted for DFT+U calculations that seem more delicate to converge self-consistently.

\subsubsection{Li-ion and magnetic configurations}
As anticipated at the beginning of this section, the same magnetic configurations considered 
for the Fe-based system are also compared for LMPO. The total energies have the
same ordering as in LFPO (E$_{AF_1}$ $<$ E$_{AF_2}$ $<$ E$_{AF_3}$ $<$ E$_{FM}$) 
with the ground state also in the AF$_1$ configuration. 
The energy differences between these different magnetic orders and the AF$_1$ ground state show
a moderate dependence on the approximation being adopted (while the ordering is robust).
In particular, the energy differences are (in meV per formula unit): 2.2, 12.3, 15.5 from DFT+U+V and 4.7, 35.4 and 43.4 in 
GGA (PBEsol) calculations. In analogy with LiFePO$_4$ these energy differences 
are irrelevant for the evaluation of formation energies and average voltages.
Their small value (actually comparable with the precision of our calculation) suggest that all these magnetic configurations
can be reached even at moderate temperatures.
\begin{figure}[h!]
\begin{center}
\subfigure[]{
\includegraphics[width=3.80cm,keepaspectratio,angle=-0]{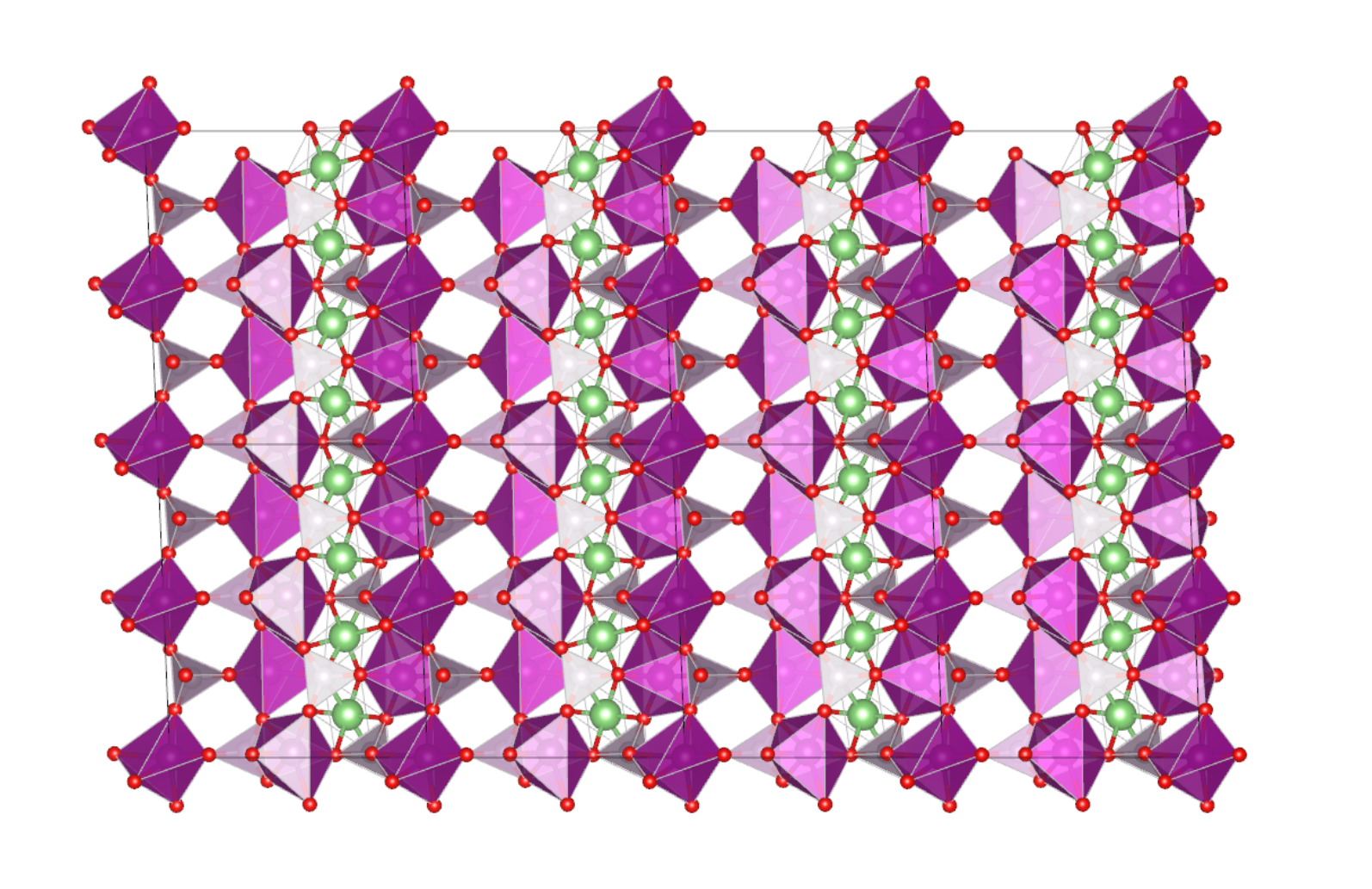}}
\subfigure[]{
\includegraphics[width=3.80cm,keepaspectratio,angle=-0]{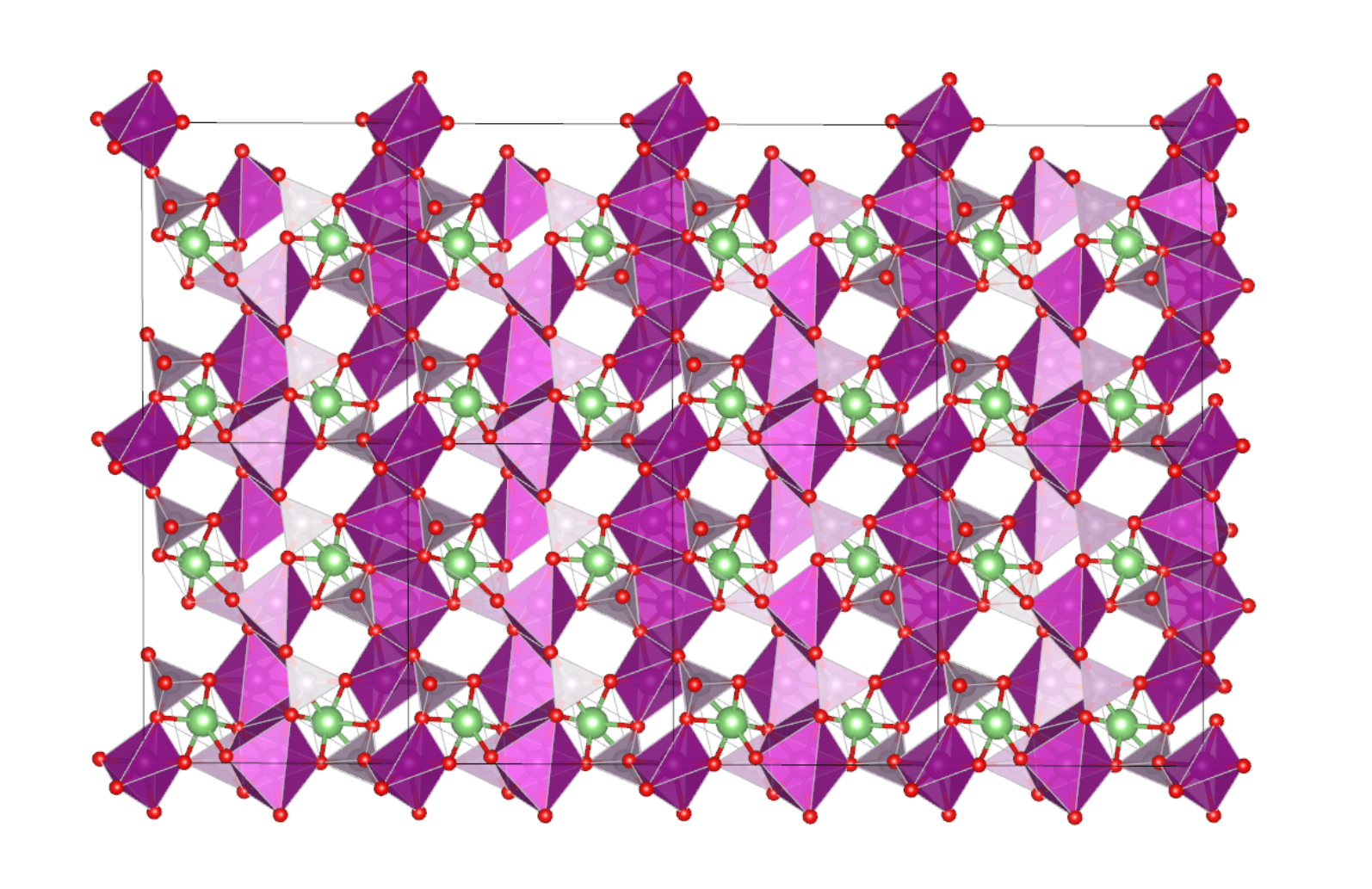}}
\subfigure[]{
\includegraphics[width=3.90cm,keepaspectratio,angle=-0]{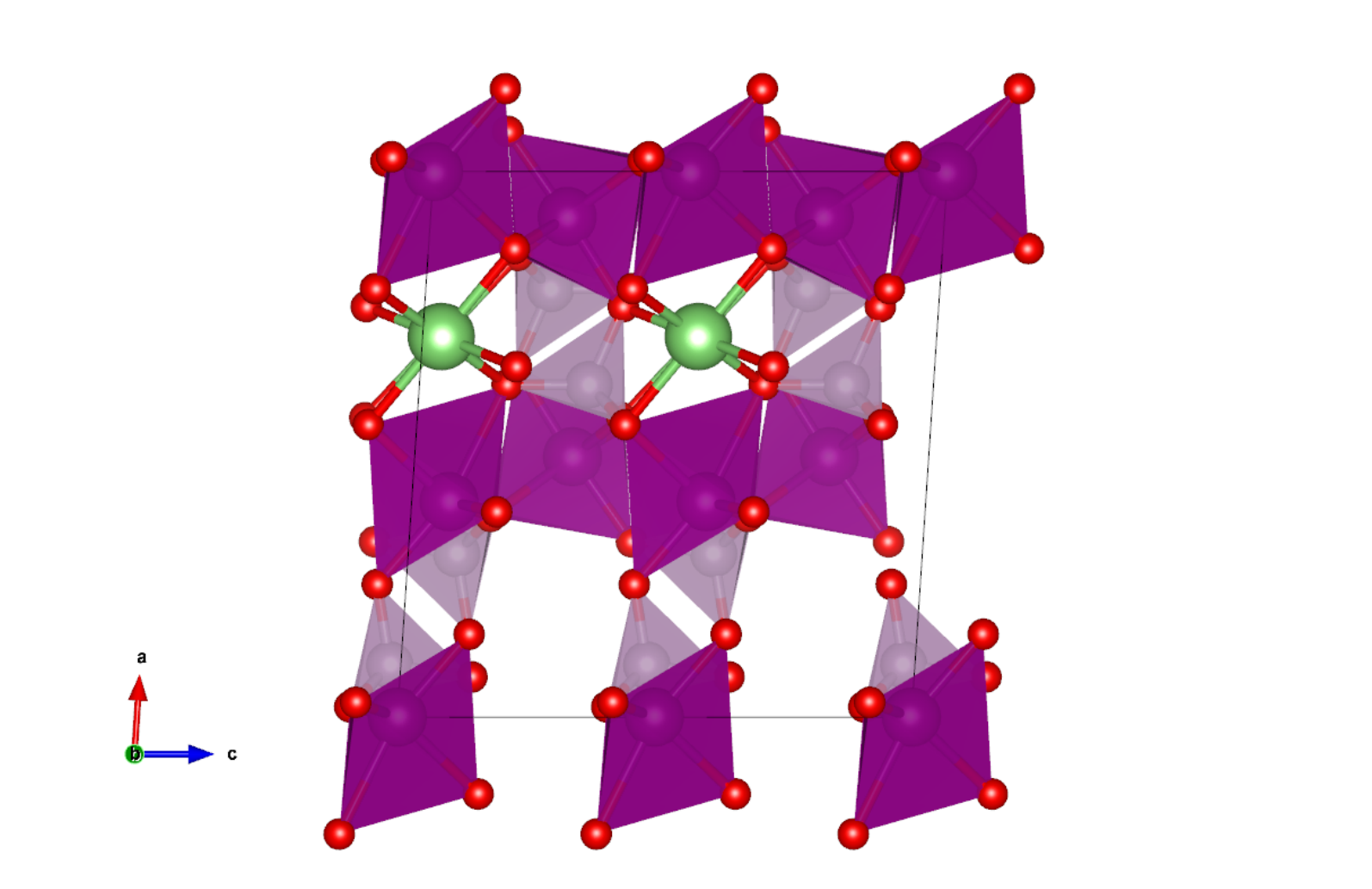}}
\subfigure[]{
\includegraphics[width=3.90cm,keepaspectratio,angle=-0]{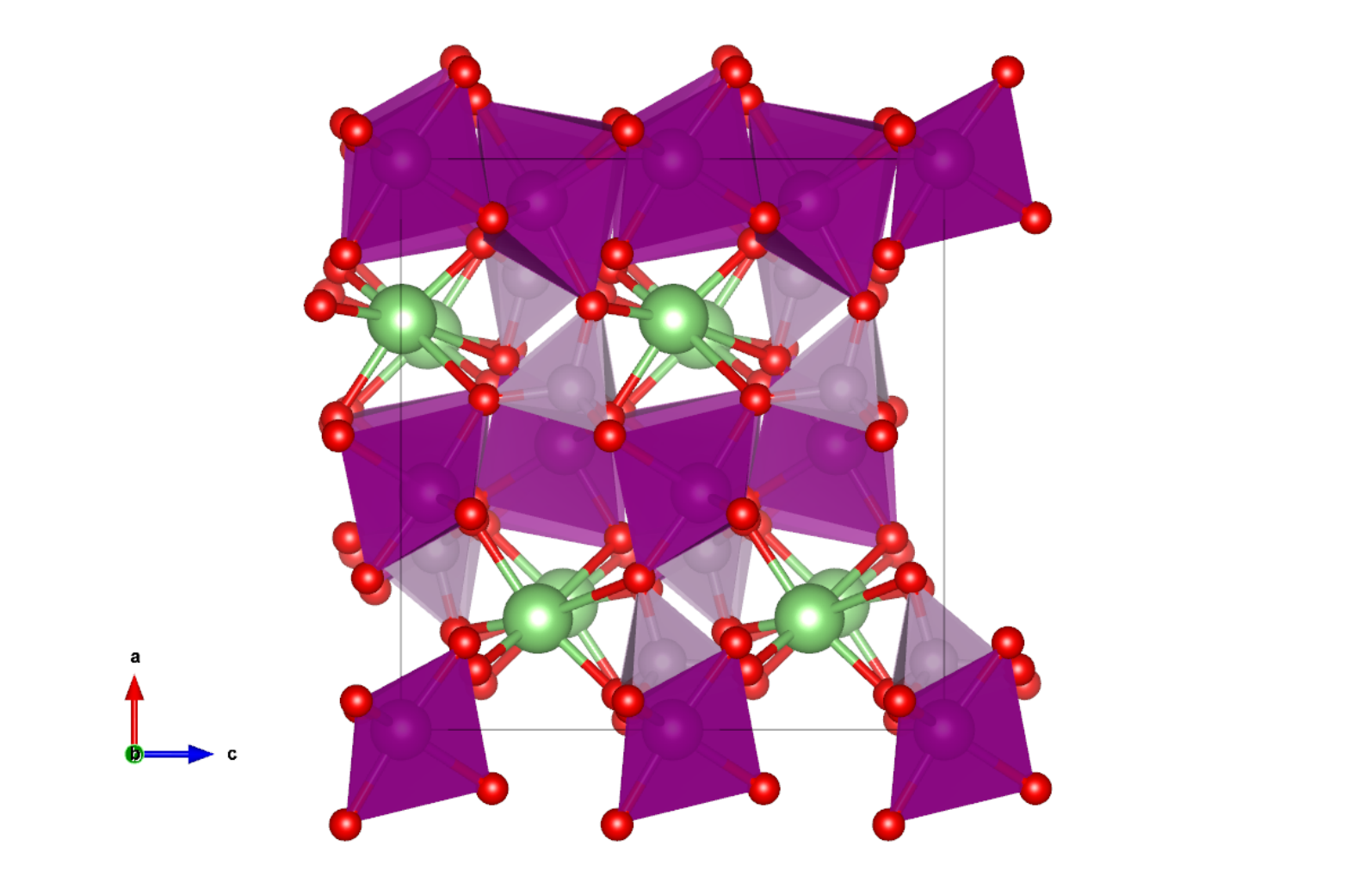}}
\caption{\label{l05mpo_12} The two Li configurations of Li$_{0.5}$MnPO$_4$ considered in this work with
Li ions occupying alternate $yz$ planes (a and c) or $xz$ planes (b and d). The figure compares views of
a 4$\times$4$\times$2 supercell of the crystals along the [001] (a and b) and along the [010] (c and d) directions.}
\end{center}
\end{figure}

Of greater interest is the comparison between the two Li configurations for the $x$ = 0.5 compound.
Based on the clear improvement brought about by the extended DFT+U+V scheme compared to other Hubbard flavors, 
as evident from the results presented in the previous section, calculations were performed only with this functional
and with uncorrected GGA for comparison.
As in the case of LiFePO$_4$, the configuration with Li filling alternate $xz$ planes has lower energy 
than the one with Li on alternate $yz$ planes and results about 255 (105) meV per formula unit more 
stable with DFT+U+V (GGA). 
At the same time the formation energy of the $xz$ structure thus amounts to 58 (-42) meV per formula unit
with DFT+U+V (GGA), which is of the same order
of magnitude of the analogous structure of the Fe olivine.
The rearrangement of the crystal structure is, however, more pronounced than in the case of the 
Fe olivine, consistently with the fact that the Mn$^{3+}$ ions are Jahn-Teller active and typically
induce more pronounced distortions to the octahedral oxygen cages around them.
Figure \ref{l05mpo_12} compares the unit cells of 
Li$_{0.5}$MnPO$_4$ with Li ions on alternate $yz$ [Figs. \ref{l05mpo_12}(a) and \ref{l05mpo_12}(c)]
and $xz$ planes [[Figs. \ref{l05mpo_12}(b) and \ref{l05mpo_12}(d)]. 
Consistently with the case of LFPO these cells will be 
referred to as $yz$ and $xz$ cells. 
As can be seen already from a visual comparison between Fig. \ref{l05mpo_12}(a) and \ref{l05mpo_12}(b), 
the $xz$ cell regularizes its shape and almost completely
recovers an orthorhombic symmetry. In fact, its equilibrium lattice parameters (bohrs) and angles (degrees)
are: $a = 19.20$, $b = 11.59$, $c = 9.15$, $\alpha = \beta = 90.0$, and $\gamma = 90.36$. 
At the same time, when Li atoms fully occupy alternate $xz$ planes, the system undergoes 
a significant distortion of its internal coordinates. 
A particularly relevant aspect of this distortion
is the fact that Li cations loose their alignment along the [010] crystalline direction and configure themselves 
in a zig-zag pattern (with more complicated arrangements possibly emerging from bigger supercells of larger extension
along $b$).
This fact is quite evident from figures \ref{l05mpo_12}(c) and \ref{l05mpo_12}(d) that, in fact, offer a view of the unit cell 
normal to the $xz$ planes (i.e., along the [010] direction).
Based on the consolidated idea that Li motion takes place along the [010] direction in olivine
phosphates \cite{leoni11,morgan04}, the observation above could provide an important element to explain 
the low ionic conductivity of the Mn compound (especially in comparison with the Fe olivine). 
In fact, the diffusion of Li ions during the charge and discharge transients moves the interface between the
$x = 1$ and $x = 0$ regions of the cathode. If a $x = 0.5$ phase is stabilized at this interface 
(e.g., by a reduction of the interfacial stress between the two phases), Li ions have to diffuse through the 
half-lithiated crystal.
While the $xz$ half-lithiated structure would be
favored by its lower formation energy and a higher degree of structural compatibility with the
$x = 0, 1$ crystals (both orthorhombic), the zig-zag ordering of the Li ions and, in general, a more distorted internal 
structure, could substantially impair the diffusion of these ions 
along the preferred [010] direction 
and thus compromise the ionic conductivity of the electrode or the capability of the boundary to
move.

The structural rearrangement that is observed between the two Li configurations also reflects a different
distribution of valence electrons (a different spatial configuration of 2+ and 3+ Mn cations). However, as in the case of
the Fe olivine, the Mn ions that result closer to the Li atoms consistently assume a 2+ state, while 
the others result in a 3+ state. Their total $d$ states occupations, approximately equal to 5.21 and 4.97
are, again, consistent with those of the $yz$ configuration for Mn ions in corresponding oxidation 
states and shown in Tables \ref{fevmn} and \ref{SI.fevmn} for DFT+U+V and DFT+U+V$_{1\times 2\times 2}$
(5.22/4.99 and 5.21/4.98, respectively). 
The values of the (self-consistent) Hubbard $U$ obtained for the Mn 2+ and 3+ species are,
respectively, 4.71 and 6.70 eV, in quite good agreement with the ones computed for the $yz$ 
structure and presented in Tables \ref{uevmn} and \ref{SI.uevmn}. 
The values of the Hubbard $V$ (0.42 - 1.35 and 0.28 - 1.03 eV, respectively) show instead a 
much looser resemblance with those in the same tables, probably due to difference in the local 
crystal structure
around Mn ions in corresponding states in the two configurations. In fact we also note
that while distances between Mn and their neighbor O ions are quite similar 
(except for the most distant two oxygens)
for the transition-metal ions in 3+ states, for the 2+ ions the oxygen cages are somewhat different. 
In addition, in the $xz$ configuration Mn ions are always closer to the nearest Li, which seems to suggest
a higher binding energy between Li ions and moving electrons, also corroborating the 
idea that the $xz$ configuration
is characterized by a low (ionic and electronic) conductivity.

\section{Summary}

In this work we have used an extended formulation of DFT+U, named DFT+U+V \cite{campo10},
to perform first-principles calculations on two important transition-metal olivine phosphates, 
Li$_x$FePO$_4$ and Li$_x$MnPO$_4$, that are studied 
as materials for cathodes of Li-ion batteries.
The new computational tool employed in these calculations is characterized by a corrective functional 
based on an extended Hubbard Hamiltonian that contains both on-site ($U$) and inter-site ($V$) 
effective interactions.
These electronic interaction parameters, computed for all the TM species (crystallographic site, magnetic and oxidation
state) present, are calculated from first-principles using linear-response theory 
\cite{cococcioni05}
and, through a recursive self-consistent procedure, are obtained in full consistency with both the electronic and
crystal structures.
Computing the energetics of the above-mentioned materials at various Li contents and evaluating
key quantities to assess their performance as the average voltage, 
this work develops a thorough comparison between the results obtained from DFT+U+V, and those 
of standard DFT functionals, such as PBEsol, or on-site only Hubbard corrections (DFT+U), 
using either an effective Hubbard $U$ computed for each Li concentration, 
or averaged over the entire chemical composition
range, as often done in literature. 
The comparative analysis is completed by contrasting the computed quantities with
experimental data, when available.

The results demonstrate that $i)$ material- and site-specific interaction parameters can be used confidently
(provided they are evaluated self-consistently) in energy comparisons without any need of averaging over 
composition ranges; $ii)$ the addition of the inter-site interaction ($V$) 
to the corrective Hubbard functionals improves significantly the accuracy of the approach and extends considerably 
the range of electronic localization regimes that can be successfully described. 
For Li$_x$FePO$_4$, the more ionic of the two systems, the improvement that this extension brings is mostly 
quantitative: it corrects the width
of the band gap of the material, refines the equilibrium crystal structure and, most notably, 
improves the value of the average voltages, delivering all these quantities
in better agreement with available experimental data. 
For Li$_x$MnPO$_4$, the inclusion of the inter-site coupling leads also to a qualitative improvement
on the results of simpler approximations. In fact, this material features a more pronounced hybridization 
between the transition metal ions and their nearest neighbor oxygens and a Hubbard corrective functional 
with on-site only interactions (i.e. DFT+U) is not capable to capture electronic localization 
(e.g., on the Mn$^{2+}$ ions at $x = 0.5$) and overcorrects 
important quantities such as the average voltage. 
The inclusion of the inter-site interaction $V$ leads to a more refined description of 
hybridized states that captures charge disproportionation at intermediate Li content (i.e., a 
clear distinction of Mn ions in 2+ and 3+ with occupations separately similar to those of lithiated 
and delithiated materials), refines the equilibrium crystal structure and the agreement with
the lattice parameters measured in experiments (albeit not uniformly for all compositions), and 
improves the prediction of the average voltage vs Li/Li$^+$. The use of the extended Hubbard functional 
also allows to capture a significant reorganization of the internal structure of the crystal upon 
changing Li configuration at intermediate compositions (probably a consequence of the Jahn-Teller 
activity of Mn$^{3+}$ ions) that suggests a stronger binding than in the Fe olivine between 
Li$^{+}$ and incoming electrons and higher kinetic barriers for Li ions to overcome
during their diffusion.
To further test the predictivity of the extended DFT+U+V functional we have also performed 
calculations on the mixed Fe-Mn olivine phosphate Li$_x$Mn$_y$Fe$_{1-y}$PO$_4$ 
(see Ref. \cite{deng17} for a useful review on this system).
Preliminary calculations (whose results will be published elsewhere) on this
system have shown that DFT+U+V is also able to capture charge disproportionation in presence of multiple
transition-metal species with occupations that are consistent with those shown in this paper for both
Fe and Mn in all the oxidation states explored. In addition it correctly predicts Mn$^{3+}$ ions in the delithiated compound
to be the first to reduce upon Li intercalation. 

From a methodological point of view the accuracy and predictivity that the extended DFT+U+V functional achieves
with a consistent calculation of the Hubbard parameters is quite remarkable, 
especially in comparison with the very abundant literature where the simpler on-site only DFT+U is used
with a semiempirical tuning of the interaction parameters on the properties of interest.
Notwithstanding the fact that the semiempirical evaluation of interaction parameters
is impossible without reference (e.g., experimental results to fit) and is far less viable
with more advanced functionals (e.g., including several kinds of interactions) or when 
multiple Hubbard species (or oxidation states) are present, it should be also kept in mind that 
within DFT+U it is often impossible to identify a single set of Hubbard parameters
able to improve the prediction of all the properties of a system \cite{reboredo14,wolverton14,sato17}. 
In other words if a value of $U$ is 
needed to reproduce the equilibrium crystal structure, the one(s) necessary to improve for example the prediction
of the magnetic moment, of the band gap, or of the energetics of certain processes (as Li intercalation)
is(are) likely to be quite different.
We believe that the results of this work show quite clearly that DFT+U+V with Hubbard parameters
evaluated self-consistently from LRT represents a significant step forward in this respect
as not only allows to confidently approach the study of a much broader spectrum of different materials,
but is also capable to improve the prediction of several properties at the same time.
The recent automation of the LRT evaluation of the Hubbard parameters through DFPT
\cite{timrov18}, reducing significantly 
its computational cost and improving its robustness, user-friendliness and accuracy is thus making
self-consistent DFT+U+V emerge as a predictive, versatile and efficient tool for the accurate modeling 
of a broad variety of materials (especially when their functionality is related to electronic localization) 
and for the discovery and optimization of new ones through large-scale high-throughput ab initio calculations.

\section{Acknowledgements}

Partial financial support for this work was provided by the Max Planck - EPFL center for Molecular Nanoscience
and Technology, by the Swiss National Science Foundation through Grant No. 200021 - 179138
and its National Centre of Competence in Research (NCCR)
MARVEL, and from the EU-H2020 research and innovation programme under grant agreement
No 654360 within the framework of the NFFA-Europe Transnational Access Activity.
We are grateful to Profs. J. Maier and B. Lotsch for suggesting this as an important case study.
The calculations presented in this paper were performed using computational resources made available
by the Swiss National Supercomputing Centre (CSCS), through Grants s580 and s836. 
Figures containing crystal structures (Figs. \ref{lfpostruc}, \ref{afmconf}, \ref{l05fpo_12}, \ref{l05mpo_12})
were all realized using the VESTA crystal visualization software \cite{vesta11}.

\bibliographystyle{unsrt}
\bibliography{refs} 

\newpage

\end{document}


\title{
\small{Supplementary information for}\\
Energetics and cathode voltages of LiMPO$_4$ olivines (M = Fe, Mn)
from extended Hubbard functionals
}

\author{Matteo Cococcioni}

\author{Nicola Marzari}

\affiliation{Theory and Simulation of Materials (THEOS), and
National Centre for Computational Design and Discovery of
Novel Materials (MARVEL),
\'Ecole Polytechnique F\'ed\'erale de Lausanne (EPFL), 
CH-1015 Lausanne, Switzerland}

\date{\today}

\maketitle

In this Supplementary Information addition we discuss the results obtained when using slightly
different strategies to compute the effective Hubbard parameters than those presented in the main paper.
Specifically, we present DFT+U+V calculations performed with a finite Hubbard $U$ on the $p$ states
of oxygen ions (named DFT+U$^{dp}$+V); with the Hubbard parameters computed in a larger 
$1\times 2\times 2$ supercell (named DFT+U+V$_{1\times 2\times 2}$) and from a 
DFPT implementation of LRT \cite{timrov18} (for both DFT+U and DFT+U+V, named DFT+U$_{DFPT}$ and
DFT+U+V$_{DFPT}$, respectively).  
Designed to give equivalent results to the
supercell, real-space approach introduced in Ref. \cite{cococcioni05}, the DFPT
implementation exploits momentum-specific (monochromatic) perturbations, which allows to greatly reduce
the computational costs, to improve the scaling of the calculations, the control
on convergence, the accuracy of final results, the user-friendliness and
automation, thus making self-consistent DFT+U and DFT+U+V calculations very straightforward
and suitable for high-throughput studies. 
For the materials studied in this work, isolated
perturbations were reconstructed as summations over monochromatic perturbations 
on a regular  1$\times$2$\times$2 {\bf q}-point grid of the Brillouin zone, equivalent to
the 1$\times$2$\times$2 supercell mentioned above. DFPT-based calculations were performed
with perturbations on Fe 3$d$ and O 2$p$ states for the DFT+U+V case and on Fe 3$d$ states only for
the DFT+U case.
In all the four approaches considered in this section the Hubbard 
parameters are computed in full consistency with both the electronic and crystal structures,
as explained in the main text. Results are presented, in dedicated sections, for both 
Li$_x$FePO$_4$ and Li$_x$MnPO$_4$.

\subsection{Li$_x$FePO$_4$}
\label{slfpo}
Table \ref{uev} shows the effective Hubbard parameters for LFPO computed from self-consistent
DFT+U$^{dp}$+V, DFT+U+V$_{1\times 2 \times 2}$, DFT+U$_{DFPT}$ and DFT+U+V$_{DFPT}$ calculations, as explained
in section \ref{MP.tecdet} of the main paper. These results complement those shown in Table \ref{MP.uev} of
the main paper.
\begin{widetext}
\begin{center}
\begin{table}[h!]
\begin{tabular}{|c|c|c|c|c|}
\hline
\hline
 & Interaction & LiFePO$_4$ & \multicolumn{1}{c|}{Li$_{0.5}$FePO$_4$} & \multicolumn{1}{c|}{
FePO$_4$}  \\
\hline
\multirow{3}{*}{DFT+U$^{dp}$+V} & U$_{Fe}$ & 4.84 & 5.43 / 4.43 & 4.57  \\
& U$_{O}$ & 10.39 - 11.18 & 8.70 - 10.42 & 9.18 - 9.55 \\
& V$_{Fe-O}$ & 0.22 - 0.79 & 0.13 - 1.12 / 0.4 - 0.79 & 0.45 - 1.10 \\
\hline
\multirow{2}{*}{DFT+U+V$_{1\times 2 \times 2}$} & U$_{Fe}$ & 4.50 & 4.86 / 4.95 & 5.31\\
& V$_{Fe-O}$ & 0.15 - 0.58 & 0.09 - 0.79 / 0.32 - 0.59 & 0.40 - 0.80\\
\hline
\multirow{1}{*}{DFT+U$_{DFPT}$} & U$_{Fe}$ & 4.81 & 5.08/5.07 & 4.99\\
\hline
\multirow{2}{*}{DFT+U+V$_{DFPT}$} & U$_{Fe}$ & 5.17 & \multirow{2}{*}{n/a}  & 5.40 \\
& V$_{Fe-O}$ & 0.41 - 0.93 &  & 0.58 - 1.07 \\
\hline
\end{tabular}
\caption{The values of $U$s and $V$s (in eV) of LFPO, obtained from self-consistent
DFT+U$^{dp}$+V, DFT+U+V$_{1\times 2 \times 2}$, DFT+U$_{DFPT}$ and DFT+U+V$_{DFPT}$ calculations (see text) 
for the three Li concentrations considered.
The ranges of values reported for the $V$ parameters refer to different
O ions in the first
coordination shell, since values vary with the M-O distance.
For Li$_{0.5}$FePO$_4$ the two sets of values refer to the Fe$^{2+}$ and Fe$^{3+}$ ions,
respectively.}
\label{uev}
\end{table}
\end{center}
\end{widetext}
%
It is important to note that the values reported in Table \ref{uev} show, in general, some 
significant differences with those in Table \ref{MP.uev} in the main paper (the best agreement
is achieve between DFT+U+V and DFT+U+V$_{DFPT}$).
These results further confirm that the self-consistent screening of the electronic
interactions determine a strong dependence of the effective Hubbard parameters
both on the chemical and crystal environment of the TM ions and on the level of 
approximation in the self-consistent procedure.
The finite difference between the results from DFT+U+V$_{1\times 2 \times 2}$ and DFT+U+V$_{DFPT}$,
in principle equivalent, is probably a consequence of a non perfect convergence of the former
and of the difficulty to avoid the effects of numerical noise when using big supercells.

Table \ref{lattice_lfpo} shows the equilibrium lattice parameters in dependence
of composition for all the approaches discussed above.
%
\begin{widetext}
\begin{center}
\begin{table}[h!]
\begin{tabular}{|c|c|c|c|c|c|}
\hline
\hline
 & & \multicolumn{1}{c|}{DFT+U$^{dp}$+V} & DFT+U+V$_{1\times 2\times 2}$ & DFT+U$_{DFPT}$ & DFT+U+V$_{DFPT}$\\
\hline
\multirow{3}{*}{LiFePO$_4$}
& a & 19.43 & 19.53 & 19.57 & 19.52 \\
& b/a & 0.58 & 0.58 & 0.58 & 0.58 \\
& c/a & 0.45 & 0.45 & 0.45 & 0.45 \\
\hline
\multirow{3}{*}{Li$_{0.5}$FePO$_4$} &
a & 19.11 & 19.10 & 19.15 & \multirow{3}{*}{n/a} \\
& b/a & 0.59 & 0.59 & 0.59 & \\
& c/a & 0.47 & 0.47 & 0.47 & \\
\hline
\multirow{3}{*}{FePO$_4$} &
a &  18.58 & 18.64 & 18.69 & 18.62\\
& b/a & 0.59 & 0.59 & 0.59 & 0.59 \\
& c/a & 0.49 & 0.49 & 0.49 & 0.49 \\
\hline
\end{tabular}
\caption{The equilibrium lattice parameters (in bohr) of LFPO
obtained within the DFT+U$^{dp}$+V, DFT+U+V$_{1\times 2\times 2}$, DFT+U$_{DFPT}$ and DFT+U+V$_{DFPT}$.}
\label{lattice_lfpo}
\end{table}
\end{center}
\end{widetext}
In comparison with the results shown in Table \ref{MP.lattice_lfpo} of the main paper, the
addition of $U$ on oxygen $p$ states always leads to a contraction of the lattice 
with respect to the closest DFT+U+V approximation, which seems to worsen the agreement with the 
experimental data for LiFePO$_4$ and only slightly improve it for FePO$_4$.
This observation is somewhat in contradiction with the results of some literature where an Hubbard
correction on the $p$ states of oxygen ligands is found to refine the bonding structure
(and bond lengths) \cite{oregan18} or electronic properties \cite{acbn015}. However, these results
were obtained with simple Hubbard functionals that did not include inter-site interactions.
We interpret our findings as highlighting how the physics to be corrected is that of
3$d$ self-interaction and of Fe 3$d$ de-hybridization with O $2p$ in
DFT+U, rather than the self-interaction of 2$p$ electrons.
It is interesting to note that, as mentioned earlier, while using a larger (1$\times 2\times$2)
cell in the calculation of $U$ and $V$'s results in values for
these quantities quite different from those obtained from smaller unit cells,
the equilibrium lattice parameters predicted are almost identical
and overall in equally good agreement with experimental values, suggesting a scarce 
sensitivity of the crystal structure to small variations of the Hubbard parameters.
The results obtained from DFT+U$_{DFPT}$ and DFT+U+V$_{DFPT}$ are indeed almost identical with those 
shown in Table \ref{MP.lattice_lfpo} of the main paper, except for the $x = 0.5$ case. 
However, the significant difference in the lattice parameters for this case might be
the consequence of the swapping between 2+ and 3+ Fe ions obtained from real-space LRT DFT+U, 
(see Table \ref{MP.fev} of the main paper and the discussion thereafter) and fixed through
the use of DFPT in the self-consistent calculation of the Hubbard $U$.

Table \ref{fev} shows Fe $d$ states occupations, formation energies of Li$_x$FePO$_4$ and
average voltages (between x = 0 and x = 1) with respect to Li/Li$^+$,
computed self-consistently within the same approaches. 
%
\begin{widetext}
\begin{center}
\begin{table}[h!]
\begin{tabular}{|c|c|c|c|c|c|}
\hline
\hline
  & Fe$^{2+}$ ($x = 1$) & Fe$^{2+}$/Fe$^{3+}$ ($x=0.5$) & Fe$^{3+}$ ($x = 0$) & F. E. (meV/f.u.) & Voltage (V) \\
\hline
DFT+U$^{dp}$+V & 6.14 & 6.13/5.55 & 5.54  & -222 & 2.39 \\
\hline
DFT+U+V$_{1\times 2\times 2}$ & 6.23 & 6.22/5.77 & 5.75 & 78  & 3.66 \\
\hline
DFT+U$_{DFPT}$ & 6.21 & 6.19/5.73 & 5.71 & 105 & 3.78 \\
\hline
DFT+U+V$_{DFPT}$ & 6.23 & n/a & 5.76 & n/a & 3.55 \\
\hline
\hline
\end{tabular}
\caption{L\"owdin total occupations of Fe 3$d$ states, formation energy, and average voltage
for Li$_x$FePO$_4$ computed with DFT+U$^{dp}$+V, DFT+U+V$_{1\times 2\times 2}$, DFT+U$_{DFPT}$ and
DFPT+U+V$_{DFPT}$.} 
\label{fev}
\end{table}
\end{center}
\end{widetext}
From a comparison with the results of Table \ref{MP.fev} of the main paper, it is immediately
evident that the addition of an on-site $U$ on the oxygen $p$ states seems to deteriorate 
the quality of predictions producing a negative formation energy and a voltage far below the
experimental value.
On the other hand, in spite of the significant differences between
the interaction parameters $U$ and $V$ of the corresponding Fe ions, especially high for LiFePO$_4$
and Li$_{0.5}$FePO$_4$ (up to 0.6 eV for the $U$),
DFT+U+V$_{1\times 2\times 2}$ predicts 
occupations for the Fe ions in the three cases that are almost identical to those obtained
with the DFT+U+V (shown in Table \ref{MP.fev}) and the equilibrium
crystal structures are basically undistinguishable, as discussed earlier.
\begin{figure}
\begin{center}
\includegraphics[width=6.0cm,keepaspectratio,angle=-0]{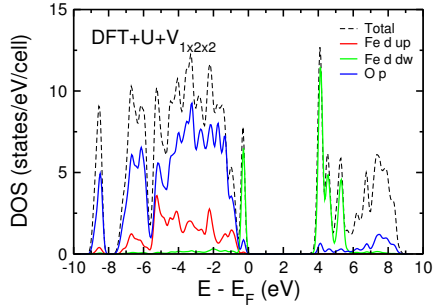}
\caption{(Color online) The density of states of LiFePO$_4$ obtained 
with DFT+U+V$_{1\times 2\times 2}$. 
The black dashed line represents the total density of state while
solid red, green and blue ones designate iron $d$ state spin up, iron $d$ state spin down and
oxygen $p$ states total contributions, respectively. All energies are referred to the Fermi level or
to the top of the valence band in presence of a gap.}
\label{lfpodos}
\end{center}
\end{figure}
However, the average voltage shows some dependence
on the values of the effective interactions, and the one
computed from DFT+U+V$_{1\times 2\times 2}$ calculations is 0.15 V higher,
in slightly worse agreement with experimental results. 
As for calculations relying on DFPT for the evaluation of Hubbard parameters, it is easy to 
check the substantial equivalence with the homologous results shown in Table \ref{MP.fev}
of the main paper. However, as mentioned earlier, the DFT+U now predicts Fe$^{2+}$ closer
to the Li ions of the $x = 0.5$ material, in consistency with chemical intuition. This
result confirms that the DFT+U results presented in the main paper actually correspond
to a local minimum of the energy.
While slightly different than the corresponding values of Table \ref{MP.fev}, the voltages 
predicted by DFT+U$_{DFPT}$ and DFT+U+V$_{DFPT}$ show a comparable agreement with experimental
results. The comparison of the DFT+U+V$_{DFPT}$ voltage with the one from DFT+U+V$_{1\times 2\times 2}$
(in principle equivalent) is instead less favorable and probably confirms the
difficulty to achieve good levels of convergence when using supercells.

Figure \ref{lfpodos} shows the projected density of states of LiFePO$_4$ 
obtained within DFT+U+V$_{1\times 2\times 2}$.
From the comparison with Fig. \ref{MP.lfpodos} of the main paper it is evident that the difference with
the DFT+U+V DOS is merely quantitative with DFT+U+V$_{1\times 2\times 2}$ resulting in a smaller
band gap (3.6 eV), actually closer to the experimental value ($\approx$ 3.7 eV). 
This observation is actually important from a methodological point of view:
while converging (e.g., with the size of the supercell)
the value of the effective electronic interaction parameters
used in the corrective functional is obviously important to capture quantities that are related
to various features of the KS spectrum, other properties that
can be computed from total energy differences seem more robust with respect
to the value of the effective interactions and show a faster convergence with size.
However, this is only true for DFT+U+V, probably due to a sort of compensation between
the variation of $V$ and that of $U$.

\subsection{Li$_x$MnPO$_4$}
\label{slmpo}
Table \ref{uevmn} shows the effective Hubbard parameters for LMPO computed from self-consistent
DFT+U$^{dp}$+V, DFT+U+V$_{1\times 2 \times 2}$, DFT+U$_{DFPT}$ and DFT+U+V$_{DFPT}$ calculations.
These results complement those shown in Table \ref{MP.uevmn} of the main paper.
\begin{widetext}
\begin{center}
\begin{table}[h!]
\begin{tabular}{|c|c|c|c|c|}
\hline
\hline
 & & LiMnPO$_4$ & \multicolumn{1}{c|}{Li$_{0.5}$MnPO$_4$} & \multicolumn{1}{c|}{MnPO$_4$}  \\
\hline
\multirow{3}{*}{DFT+U$^{dp}$+V} & U$_{Mn}$ & 3.84 & \multirow{3}{*}{n/a} & 5.52\\
& U$_{O}$ & 10.0 - 11.29 &  & 8.38 - 8.45\\
& V$_{Mn-O}$ & 0.45 - 1.03 &  & 0.33 - 1.25\\
\hline
\multirow{2}{*}{DFT+U+V$_{1\times 2\times 2}$} & U$_{Mn}$ & 3.67  &  4.60 / 6.80 & 5.88 \\
& V$_{Mn-O}$ & 0.38 - 0.71  & 0.17 - 1.05 / 0.56 - 1.17  & 0.31 - 0.96 \\
\hline
\multirow{1}{*}{DFT+U$_{DFPT}$} & U$_{Mn}$ & 3.78 & 7.85 / 8.93 & 6.12  \\
\hline
\multirow{2}{*}{DFT+U+V$_{DFPT}$} & U$_{Mn}$ & 4.22  &  \multirow{2}{*}{n/a} & 6.42 \\
& V$_{Mn-O}$ & 0.67 - 1.08  &  & 0.59 - 1.37 \\
\hline
\end{tabular}
\caption{The values of $U$s and $V$s (in eV) of LMPO for the three considered Li concentrations
computed within DFT+U$^{dp}$+V, DFT+U+V$_{1\times 2\times 2}$, DFT+U$_{DFPT}$ and DFT+U+V$_{DFPT}$.
The ranges of values reported for the $V$ parameters refer to the first
coordination shell (their values vary with the M-O distance).
For Li$_{0.5}$MnPO$_4$ the two sets of values refer to the 2+ nd 3+ Mn ions
respectively.}
\label{uevmn}
\end{table}
\end{center}
\end{widetext}
%
Based on the results obtained for LFPO self-consistent calculations of
$U$'s and $V$'s were not performed within DFT+U$^{dp}$+V or within DFT+U+V$_{DFPT}$ 
for the half lithiated material (due to higher structural freedom and
lower symmetry, calculations for this specific system are typically more expensive and harder to converge).
This omission only precludes the possibility to determine for this functional the formation energy
for $x = 0.5$ and to evaluate
the occupation of the two classes of Mn ions in comparison with the end-members of the family; the average
voltage, instead, depending only on the comparison between the $x=0$ and $x=1$ materials, can be easily
evaluated and will be discussed later.
The difference between the values obtained with DFT+U$^{dp}$+V and those reported in Table \ref{MP.uevmn}
(especially for DFT+U+V, the closest approximation) seems to indicate an important role of oxygen $p$
states in the screening of the effective interactions; at the same time the comparison between 
DFT+U+V$_{1\times 2\times 2}$ and DFT+U+V suggests a relatively slow convergence of the Hubbard parameters
with the size of the supercell used in their calculation. 
\begin{widetext}
\begin{center}
\vspace{0mm}
\begin{table}
\begin{tabular}{|c|c|c|c|c|c|}\hline
\hline
 & & \multicolumn{1}{c|}{DFT+U$^{dp}$+V} &  DFT+U+V$_{1\times 2\times 2}$ & DFT+U$_{DFPT}$ & DFT+U+V$_{DFPT}$ \\
\hline
\multirow{3}{*}{LiMnPO$_4$} &
 a & 19.70  & 19.79 & 19.86 & 19.78\\
& b/a & 0.58 & 0.58 & 0.58 & 0.58 \\
& c/a & 0.45 & 0.45 & 0.45 & 0.45 \\
\hline
\multirow{6}{*}{Li$_{0.5}$MnPO$_4$} &
a & \multirow{6}{*}{n/a} & 19.17 & 19.46 & \multirow{6}{*}{n/a} \\
& b/a & & 0.61 & 0.60 & \\
& c/a & & 0.47 & 0.47 & \\
& $\alpha$ & & 89.77  & 90.0 & \\
& $\beta$ & & 86.21 & 87.09 & \\
& $\gamma$ & & 92.29 & 90.0 & \\
\hline
\multirow{3}{*}{MnPO$_4$} &
a & 19.25 & 18.73 & 18.72 & 18.76 \\
& b/a & 0.54 & 0.61  & 0.61 & 0.61 \\
& c/a & 0.52 & 0.50 & 0.50 & 0.50 \\
\hline
\end{tabular}
\caption{The equilibrium lattice parameters (in bohr) and the angles between the primitive cell vectors 
of LMPO obtained with DFT+U$^{dp}$+V, DFT+U+V$_{1\times 2\times 2}$, DFT+U$_{DFPT}$ and DFT+U+V$_{DFPT}$.}
\label{lpmn}
\end{table}
\end{center}
\end{widetext}
%
Regarding the results of calculations based on DFPT it is easy to realize that there is a quite significant
discrepancy with those in Table \ref{MP.uevmn} of the main paper for DFT+U. However, for DFT+U+V the converged
values for $U$ and $V$ agree quite nicely with those reported in Table \ref{MP.uevmn}, both for $x = 0$ and $x = 1$.
This agreement is actually of better quality than the comparison with DFT+U+V$_{1\times 2\times 2}$, to which 
DFT+U+V$_{DFPT}$ is in principle equivalent. 
This observation further confirm the difficulty in controlling the numerical
convergence of finite difference calculations performed in big supercell and the advantages of the DFPT implementation.
While these considerations on the numerical values of the Hubbard parameters are practically useful, it is 
important to remark that $i)$ they contain the effects
of structural optimization, which is part of the self-consistent procedure; $ii)$ a more meaningful
and important comparison is to be developed on the results obtained from calculations performed
within the various functionals. This analysis will be expanded in the reminder of this section.

Table \ref{lpmn} shows the equilibrium lattice parameters in dependence
of composition for DFT+U$^{dp}$+V, DFT+U+V$_{1\times 2 \times 2}$, DFT+U$_{DFPT}$ and DFT+U+V$_{DFPT}$. These results are to
be compared with those in Table \ref{MP.lpmn} of the main paper.
As in the case of LFPO, it is worth noting that, in spite of the differences
between the values of $U$'s and $V$'s computed in the single or 1$\times$2$\times$2
supercell, the equilibrium lattice parameters predicted in the two cases
are practically coincident for all three Li concentrations. 
With DFT+U+V$_{1\times 2\times 2}$ however the angles between
the unit cell vectors of Li$_{0.5}$MnPO$_4$ are slightly closer to 90$^{\circ}$ than with DFT+U+V.
The addition of a finite Hubbard $U$ on the $p$ state of oxygen through DFT+U$^{dp}$+V
contracts the lattice parameter for x = 1 (resulting in an agreement with experimental data
of the same quality of DFT+U+V) but predicts a lattice for the x = 0 crystal significantly 
expanded with respect to the experimental structure.
With Hubbard interactions evaluated from DFPT DFT+U+V predicts a crystal structure in very close agreement with
the corresponding ones in Table \ref{MP.lpmn} of the main paper and with results from DFT+U+V$_{1\times 2 \times 2}$.
As for DFT+U, instead, the predicted equilibrium structure is in better agreement with experiments for $x = 0$ and
$x = 1$, but determine a significant expansion of the lattice for $x = 0.5$.

Table \ref{fevmn} shows Mn $d$ states total occupations, formation energies of Li$_x$MnPO$_4$ and
average voltages (between x = 0 and x = 1) with respect to Li/Li$^+$
for all the four approaches considered above.
%
\begin{widetext}
\begin{center}
\begin{table}[h!]
\begin{tabular}{|c|c|c|c|c|c|}
\hline
\hline
  & Mn$^{2+}$ ($x = 1$) & Mn$^{2+}$/Mn$^{3+}$ ($x=0.5$) & Mn$^{3+}$ ($x = 0$) & F. E. (meV/f.u.) & Voltage (V) \\
\hline
DFT+U$^{dp}$+V & 5.15 & n/a & 4.80 & n/a & 2.52 \\
\hline
DFT+U+V$_{1\times 2\times 2}$ & 5.23 & 5.21/4.98 & 4.98 & 313 & 4.26 \\
\hline
DFT+U$_{DFPT}$ & 5.21 & 5.10/5.08 & 4.94 & 629 & 4.67  \\
\hline
DFT+U+V$_{DFPT}$ & 5.23 & n/a  & 4.98 & n/a  &  4.21 \\
\hline
\end{tabular}
\caption{Atomic occupations of $d$ states, formation energy, and average voltages
for Li$_x$MnPO$_4$, computed with DFT+U$^{dp}$+V and DFT+U+V$_{1\times 2\times 2}$.}
\label{fevmn}
\end{table}
\end{center}
\end{widetext}
%
From the comparison with the results shown in 
Table \ref{MP.fevmn} of the main paper it is easy to notice that 
the use of a Hubbard $U$ on oxygen $p$ states deteriorates the agreement of the predicted
average voltage with respect to the experimental value.
Conversely, in spite of the significant difference in the computed values of $U$ and $V$,
DFT+U+V$_{1\times 2\times 2}$ predicts an average voltage vs Li/Li$^+$ that is only slightly larger
than the DFT+U+V result (4.26 V instead of 4.15 V), still representing a good approximation
of the experimental value.
Consistently with better equilibrium structure for $x = 0$ and $x = 1$, DFT+U$_{DFPT}$ improves the 
predicted average voltage with respect to the corresponding result of Table \ref{MP.fevmn} in the main paper.
However no progress is recorded in capturing charge disproportionation for the $x = 0.5$ composition.
Actually, the significant increase in the (positive) formation energy for this material suggests
that the self-consistent procedure has probably incurred in a local minimum, with a significantly expanded 
structure.
DFT+U+V$_{DFPT}$ predicts atomic occupations in agreement with previous DFT+U+V and DFT+U+V$_{1\times 2\times 2}$
results; the average voltage it gives, 4.21 V, is actually in better agreement with DFT+U+V and experimental results
than DFT+U+V$_{1\times 2\times 2}$.

\begin{figure}
\begin{center}
\includegraphics[width=6.0cm,keepaspectratio,angle=-0]{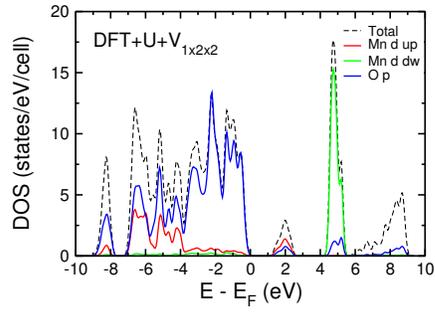}
\caption{\label{lmpodos} (Color online) The density of states of MnPO$_4$ obtained
with DFT+U+V$_{1\times 2\times 2}$.
The graphs the black dashed line represents the total density of state while
solid red, green and blue ones designate manganese $d$ state spin up, manganese $d$ state
spin down and oxygen $p$ states total contributions. All energies are referred to the Fermi
level or to the top of the valence band in presence of a gap.}
\end{center}
\end{figure}

Figure \ref{lmpodos} shows the projected density of states of MnPO$_4$
obtained within DFT+U+V$_{1\times 2\times 2}$.
From the comparison with the DOS shown in Fig. \ref{MP.lmpodos} of the main paper,
it is evident that the difference between DFT+U+V and DFT+U+V$_{1\times 2\times 2}$ is, again,
only quantitative and mostly lies on the size of
the gap that is predicted to be of about 1.5 and 1.2 eV, respectively.
As in the case of LFPO this change in the gap width is the result of the variation of
the self-consistent effective interactions and the consequent difference in
the equilibrium crystal structure. However, it is important to note that the
difference between the two DOS is quantitatively smaller than the difference between
the Hubbard interaction parameters shown in Tables \ref{uevmn} and \ref{MP.uevmn} of the
main paper. 

\bibliographystyle{unsrt}
\bibliography{refs}